\documentclass[useAMS, usenatbib]{mn2e}
\usepackage{amsmath}
\usepackage{graphicx}
\usepackage{threeparttable}
\usepackage{booktabs}

\newcommand\be{\begin{equation}}
\newcommand\ee{\end{equation}}
\newcommand\ba{\begin{eqnarray}}
\newcommand\ea{\end{eqnarray}}

\def\go{\mathrel{\raise.3ex\hbox{$>$}\mkern-14mu
             \lower0.6ex\hbox{$\sim$}}}
\def\lo{\mathrel{\raise.3ex\hbox{$<$}\mkern-14mu
             \lower0.6ex\hbox{$\sim$}}}

\title[Inertial-acoustic modes in BH Accretion Discs] 
{Simulations of Overstable Inertial-acoustic Modes in Black-Hole Accretion Discs}
\author[W.~Fu \& D.~Lai] {Wen Fu$^{1,2,3}$\thanks{Email:
    wenfu@astro.cornell.edu (WF); dong@astro.cornell.edu (DL)} and
  Dong Lai$^{1}$\footnotemark[1]\\ $^1$Department of Astronomy,
  Cornell University, Ithaca, NY 14853, USA\\ $^2$Theoretical
  Division, Los Alamos National Laboratory, Los Alamos, NM 87545,
  USA\\ $^3$Department of Physics and Astronomy, Rice University,
  Houston, TX 77005, USA}

\pagerange{\pageref{firstpage}--\pageref{lastpage}} \pubyear{2012}

\begin{document}

\label{firstpage}
\maketitle

\begin{abstract}
We present two-dimensional inviscid hydrodynamic simulations of
overstable inertial-acoustic oscillation modes (p-modes) in black-hole
accretion discs. These global spiral waves are trapped in the
inner-most region of the disc, and are driven overstable by wave
absorption at the corotation resonance ($r_c$) when the gradient of
the background disc vortensity (vorticity divided by surface density)
at $r_c$ is positive and the disc inner boundary is sufficiently
reflective. Previous linear calculations have shown that the growth
rates of these modes can be as high as $10\%$ of the rotation
frequency at the disc inner edge.  We confirm these linear growth
rates and the primary disc oscillation frequencies in our simulations
when the mode amplitude undergoes exponential growth. We show that the
mode growth saturates when the radial velocity perturbation becomes
comparable to the disc sound speed. During the saturation stage, the
primary disc oscillation frequency differs only slightly (by less than
a few percent) from the linear mode frequency. Sharp features in the 
fluid velocity profiles at this stage suggest that the saturation results
from nonlinear wave steepening and mode-mode interactions.
\end{abstract}


\section{Introduction}

In several recent papers (Lai \& Tsang 2009; Tsang \& Lai 2009c; Fu \&
Lai 2011; Horak \& Lai 2013), we have presented detailed study of the
linear instability of non-axisymmetric inertial-acoustic modes [also
called p-modes; see Kato (2001) and Wagoner (2008) for review] trapped 
in the inner-most region of black-hole (BH) accretion discs. This global instability
arises because of wave absorption at the corotation resonance (where
the wave pattern rotation frequency matches the background disc
rotation rate) and requires that the disc vortensity has a positive
gradient at the corotation radius (see Narayan et al.~1987, Tsang \& Lai 2008 
and references therein).  The disc vortensity (vorticity divided by
surface density) is given by
\be
\zeta={\kappa^2\over 2\Omega\Sigma},
\label{eq:zeta}\ee 
where $\Omega(r)$ is the disc rotation frequency, $\kappa(r)$ is the 
radial epicyclic frequency and $\Sigma(r)$ is the surface density
\footnote{Equation (\ref{eq:zeta}) applies to barotropic discs in 
Newtonian (or pseudo-Newtonina) theory. See Tsang \& Lai (2009c)
for non-barotropic discs and Horak \& Lai (2013) for full
general relativistic expression.}. 
General relativistic (GR) effect
plays an important role in the instability: For a Newtonian disc, with
$\Omega=\kappa\propto r^{-3/2}$ and relatively flat $\Sigma(r)$
profile, we have $d\zeta/dr<0$, so the corotational wave absorption
leads to mode damping. By contrast, $\kappa$ is non-monotonic near a
BH (e.g., for a Schwarzschild BH, $\kappa$ reaches a maximum at
$r=8GM/c^2$ and goes to zero at $r_{\rm ISCO}=6GM/c^2$), the
vortensity is also non-monotonic. Thus, p-modes with frequencies such
that $d\zeta/dr>0$ at the corotation resonance are overstable. Our
calculations based on several disc models and Paczynski-Witta 
pseudo-Newtonian potential (Lai \& Tsang 2009, Tsang \& Lai 2009c)
and full GR (Horak \& Lai 2013) showed that the lowest-order
p-modes with $m=2,3,4,\cdots$ have the largest growth rates, with the
mode frequencies $\omega \simeq \beta m\Omega_{\rm ISCO}$ (thus giving
commensurate frequency ratio $2:3:4,\cdots$), where the dimensionless
constant $\beta\lo 1$ depends weakly the disc properties.
These overstable p-modes could potentially explain the 
High-frequency Quasi-Periodic Oscillations (HFPQOs) observed in BH
X-ray binaries (e.g., Remillard \& McClintock 2006; Belloni et al.~2012).

The effects of magnetic fields on the oscillation modes of BH
accretion discs have been investigated by Fu \& Lai (2009, 2011, 2012)
and Yu \& Lai (2013). Fu \& Lai (2009) showed that the the basic wave
properties (e.g., propagation diagram) of p-modes are not strongly
affected by disc magnetic fields, and it is likely that these p-modes
are robust in the presence of disc turbulence (see Arras, Blaes \&
Turner 2006; Reynolds \& Miller 2009). By contrast, other diskoseismic
modes with vertical structure (such as g-modes and c-modes) may be
easily ``destroyed'' by the magnetic field (Fu \& Lai 2009) or
suffer damping due to corotation resonance (Kato 2003; 
Li et al.~2003; Tsang \& Lai 2009a). Although a modest
toroidal disc magnetic field tends to reduce the growth rate of the
p-mode (Fu \& Lai 2011), a large-scale poloidal field can enhance the instability 
(Yu \& Lai 2013; see Tagger \& Pallet 1999; Tagger \&
Varniere 2006). The p-modes are also influenced by the magnetosphere
that may exist inside the disc inner edge (Fu \& Lai 2012).

So far our published works are based on linear analysis.  While these
are useful for identifying the key physics and issues, the nonlinear
evolution and saturation of the mode growth can only be studied by
numerical simulations. It is known that fluid perturbations near the
corotation resonane are particularly prone to become nonlinear (e.g.,
Balmforth \& Korycansky 2001; Ogilvie \& Lubow 2003).  
Moreover, real accretion discs are more complex than any
semi-analytic models considered in our previous works.  Numerical MHD
simulations (including GRMHD) are playing an increasingly important
role in unraveling the nature of BH accretion flows (e.g., De Villiers
\& Hawley 2003; Machida \& Matsumoto 2003; Fragile et al.~2007; Noble
et al.~2009, 2011; Reynolds \& Miller 2009; Beckwith et al.~2008, 2009;
Moscibrodzka et al.~2009; Penna et al.~2010; Kulkarni et
al.~2011; Hawley et al.~2011; O'Neill et al.~2011; Dolence et al.~2012;
McKinney et al.~2012; Henisey et al.~2012).  Despite much
progress, global GRMHD simulations still lag far behind observations,
and so far they have not revealed clear signs of HFQPOs that are directly 
comparable with the observations of BH X-ray binaries
\footnote{Henisey et al.~(2009, 2012) found evidence of
excitation of wave modes in simulations of tilted BH accretion disks.
Hydrodynamic simulations using $\alpha$-viscosity (Chan 2009; O'Neill et al.~2009)
showed wave generation in the inner disk region by viscous instability
(Kato 2001). The MHD simulations by O'Neill et al.~(2011) revealed
possible LFQPOs due to disk dynamo cycles. 
Dolence et al.~(2012) reported transient QPOs in the numerical models
of radiatively inefficent flows for Sgr A$^\star$. McKinney et
al.~(2012) found QPO signatures associated with the interface between
the disc inflow and the bulging jet magnetosphere (see Fu \& Lai
2012). Note that the observed HFQPOs are much weaker than LFQPOs,
therefore much more difficult to obtain by brute-force simulations.}.
If the corotation instability and its magnetic counterparts studied in
our recent papers play a role in HFQPOs, the length-scale involved
would be small and a proper treatment of flow boundary conditions is
important. It is necessary to carry out ``controlled'' numerical
experiments to capture and evaluate these subtle effects.

In this paper, we use two-dimensional hydrodynamic simulations to
investigate the nonlinear evolution of corotational instability of
p-modes. Our 2D model has obvious limitations. For example it does not
include disc magnetic field and turbulence. However, we emphasize that
since the p-modes we are studying are 2D density waves with no
vertical structure, their basic radial ``shapes'' and real frequencies
may be qualitatively unaffected by the turbulence (see Arras et
al. 2006; Reynolds \& Miller 2009; Fu \& Lai 2009). Indeed, several
local simulations have indicated that density waves can propagate in
the presence of MRI turbulence (Gardiner \& Stone 2005; Fromang et
al. 2007; Heinemann \& Papaloizou 2009). Our goal here is to
investigate the saturation of overstable p-modes and the their
nonlinear behaviours.

\section{Numerical Setup}

Our accretion disc is assumed to be inviscid and geometrically thin so that the
hydrdynamical equations can be reduced to two-dimension with vertically
integrated quantities. We adopt an isothermal equation of state 
throughout this study, i.e. $P=c_s^2\Sigma$ where $P$ is the vertically integrated
pressure, $\Sigma$ is the surface density and $c_s$ is the constant
sound speed. Self-gravity and magnetic fields are neglected.

We use the Paczynski-Witta Pseudo-Newtonian potential (Paczynski \&
Witta 1980) to mimic the GR effect:
\be
\Phi=-\frac{GM}{r-r_{\rm S}},
\ee
where $r_{\rm S}=2GM/c^2$ is the Schwarzschild radius. 
The corresponding Keplerian rotation frequency and radial epicyclic frequency are
\ba
&&\Omega_{\rm K}=\sqrt{\frac{GM}{r}}\frac{1}{r-r_{\rm S}},\\
&&\kappa=\Omega_{\rm K}\sqrt{\frac{r-3r_{\rm S}}{r-r_{\rm S}}}.
\label{eq:grkappa}
\ea
In our compuation, we will adopt the units such that 
the inner disc radius (at the Inner-most Stable Circular Orbit or ISCO)
is at $r=1.0$ and the Keperian frequency at the ISCO is
$\Omega_{\rm ISCO}=1$. In these units, $r_{\rm s}=1/3$, and
\be
\Omega_{\rm K}=\frac{2}{3}\frac{1}{r-r_{\rm s}}\frac{1}{\sqrt{r}}.
\ee
Our computation domain extends from $r=1.0$ to $r=4.0$ in the radial
direction and from $\phi=0$ to $\phi=2\pi$ in the azimuthal
direction. We also use the Keplerian orbital period
($T=2\pi/\Omega_{\rm K}=2\pi$) at $r=1$ as the unit for time.
The equilibrium state of the disk is axisymmetric. The surface density
profile has a simple power-law form
\be
\Sigma_0=r^{-1},
\ee
which leads to a positive vortensity gradient in the inner disc region.
The equilibrium rotation frequency of the disc is given by 
\be
\Omega_0(r)=\sqrt{\frac{4/9}{r(r-1/3)^2}-\frac{c_s^2}{r^2}}.
\ee
Throughout our simulation, we will adopt $c_s=0.1$ so that
$\Omega_0 \simeq \Omega_{\rm K}$.

\begin{figure*}
\begin{center}
$
\begin{array}{ccc}
\includegraphics[width=0.33\textwidth]{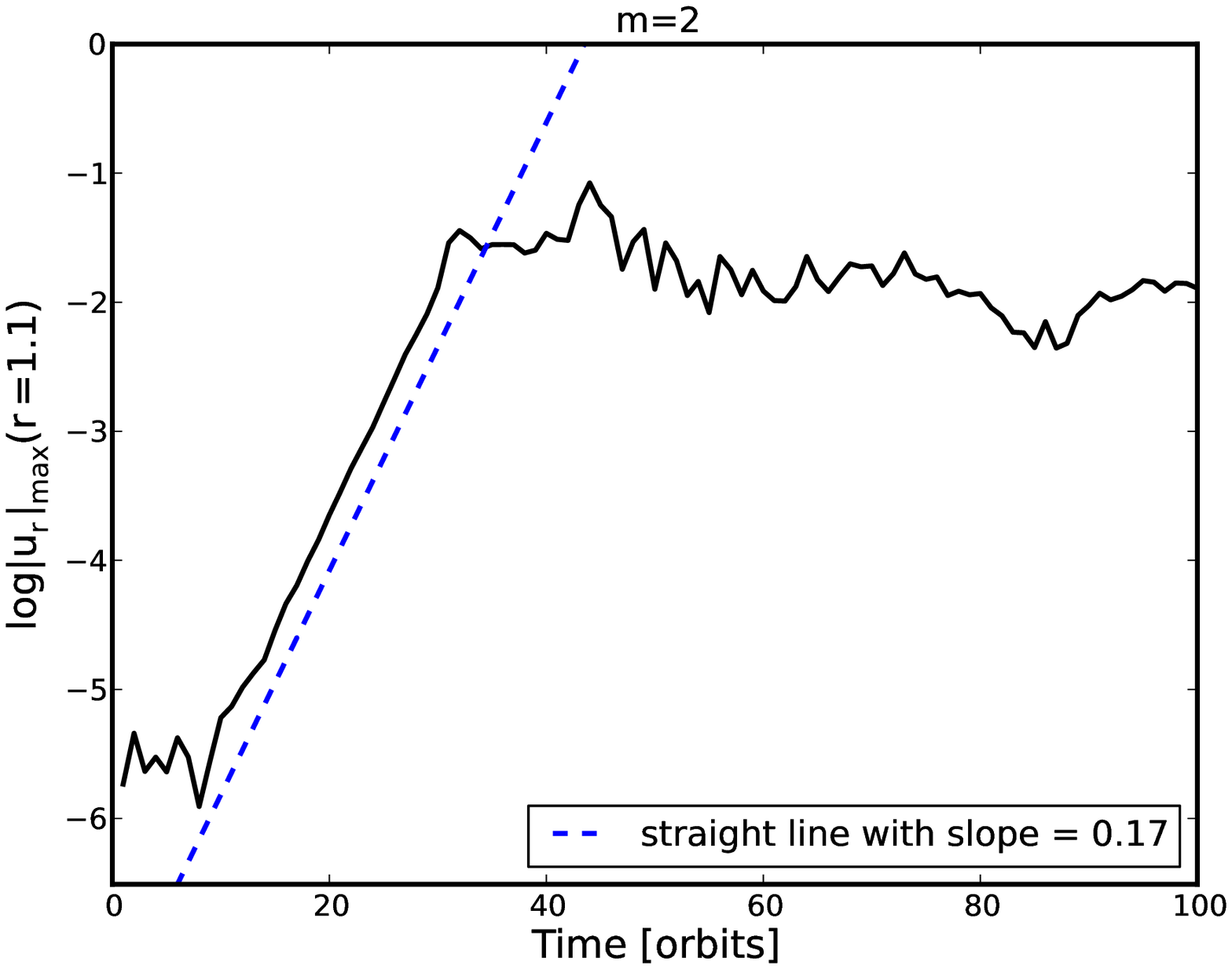} &
\includegraphics[width=0.33\textwidth]{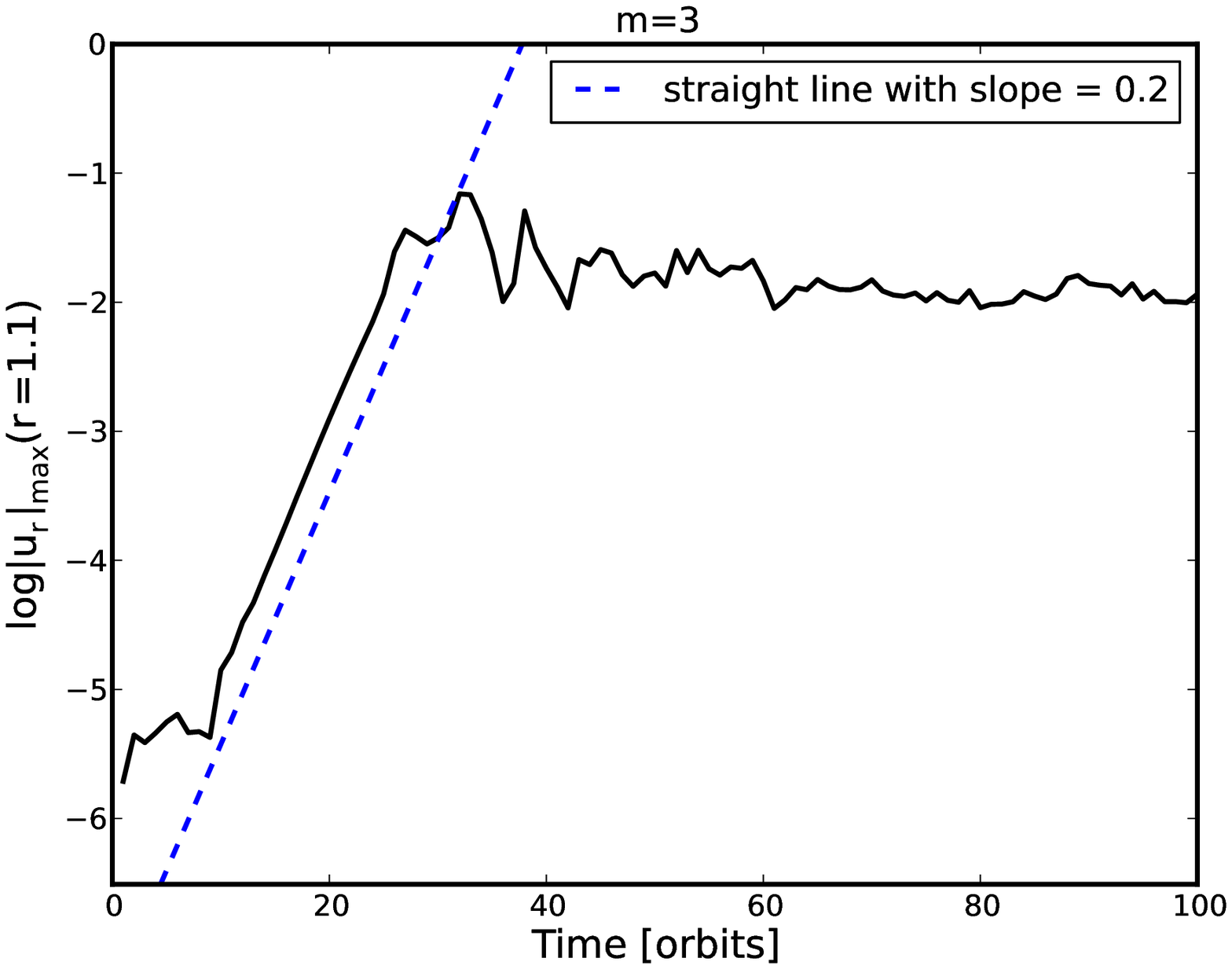} &
\includegraphics[width=0.33\textwidth]{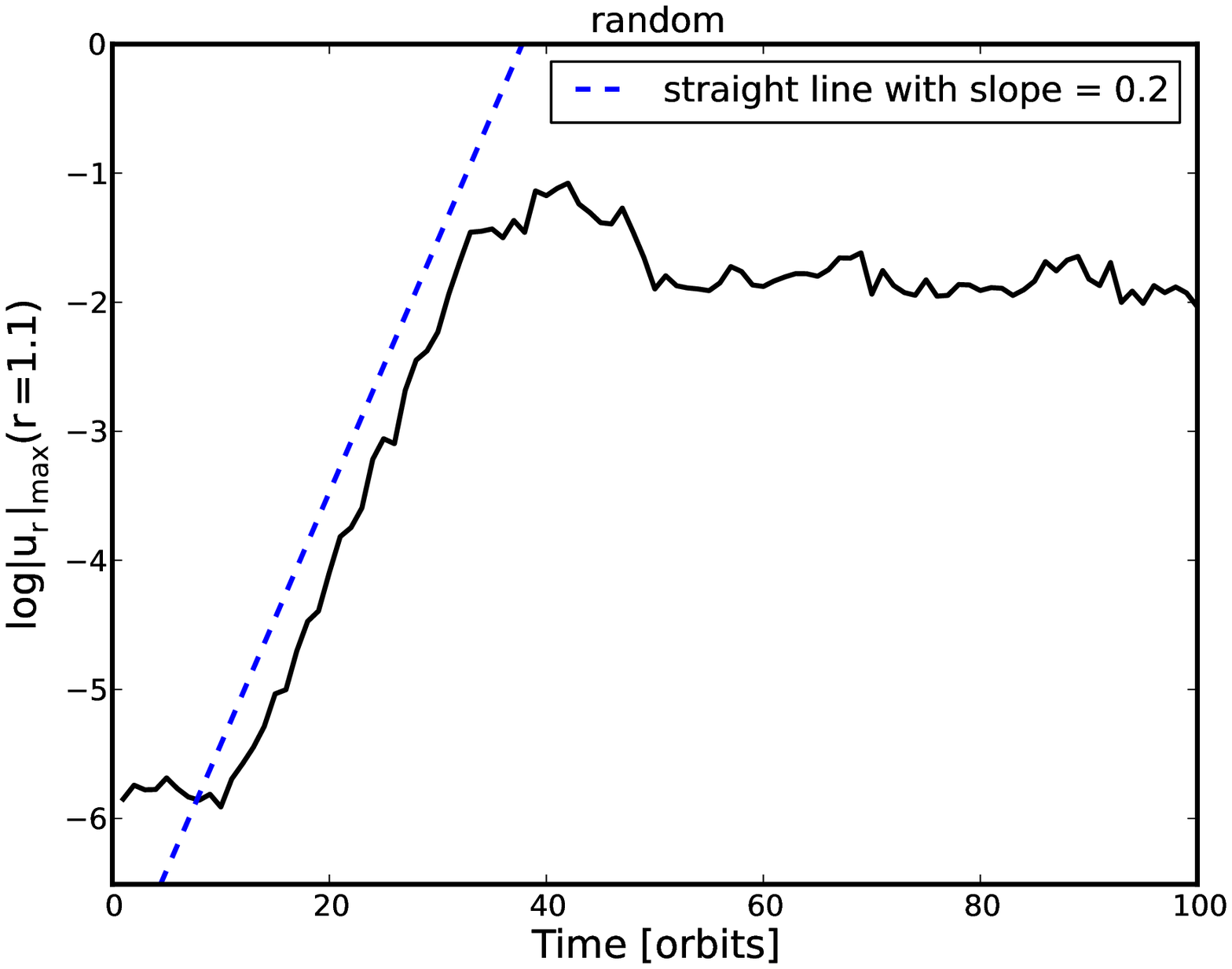}
\end{array}
$
\caption{Evolution of the radial velocity amplitude $|u_r|_{\rm max}$ 
(evaluated at $r=1.1$) for three runs with initial azimuthal mode number $m=2$
(left panel), $m=3$ (middle panel) and 
with random perturbations (right panel). 
The dashed lines are the fits for the
exponential growth stage (between $\sim 10$ and $\sim 30$ orbits) of
the mode amplitude.}
\label{fig:growth}
\end{center}
\end{figure*}

\begin{figure*}
\begin{center}
$
\begin{array}{cc}
\includegraphics[width=0.45\textwidth]{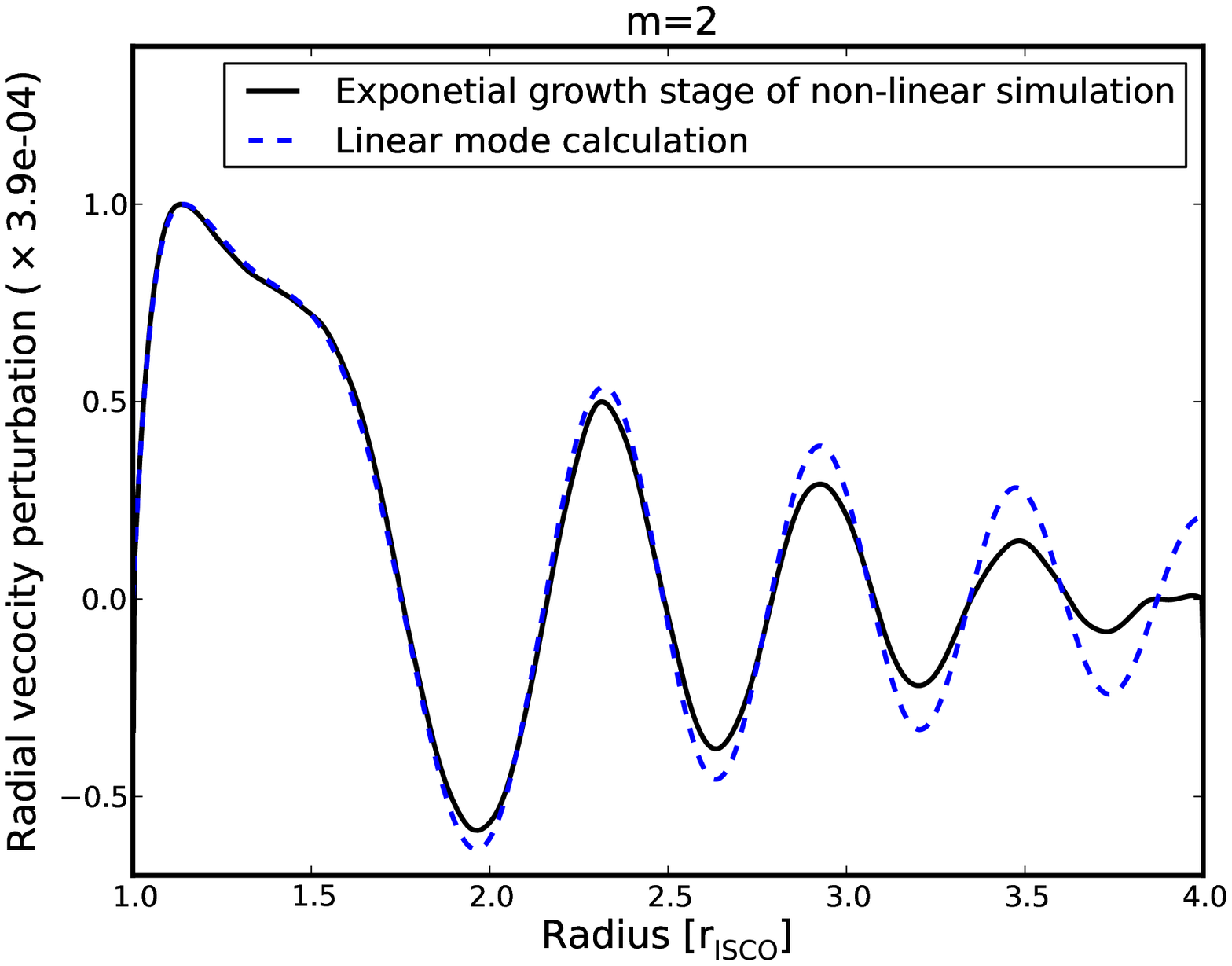} & 
\includegraphics[width=0.45\textwidth]{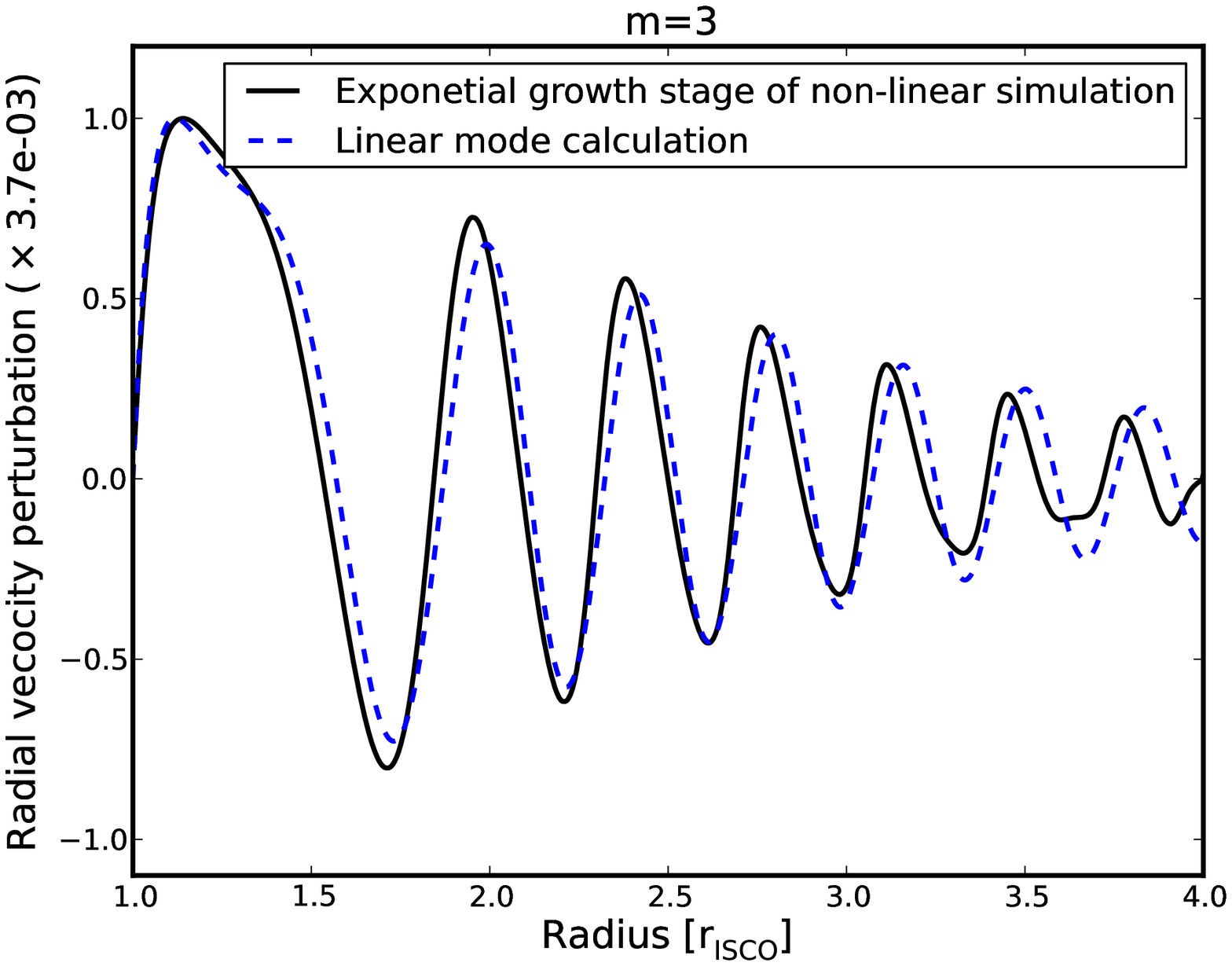} \\
\includegraphics[width=0.45\textwidth]{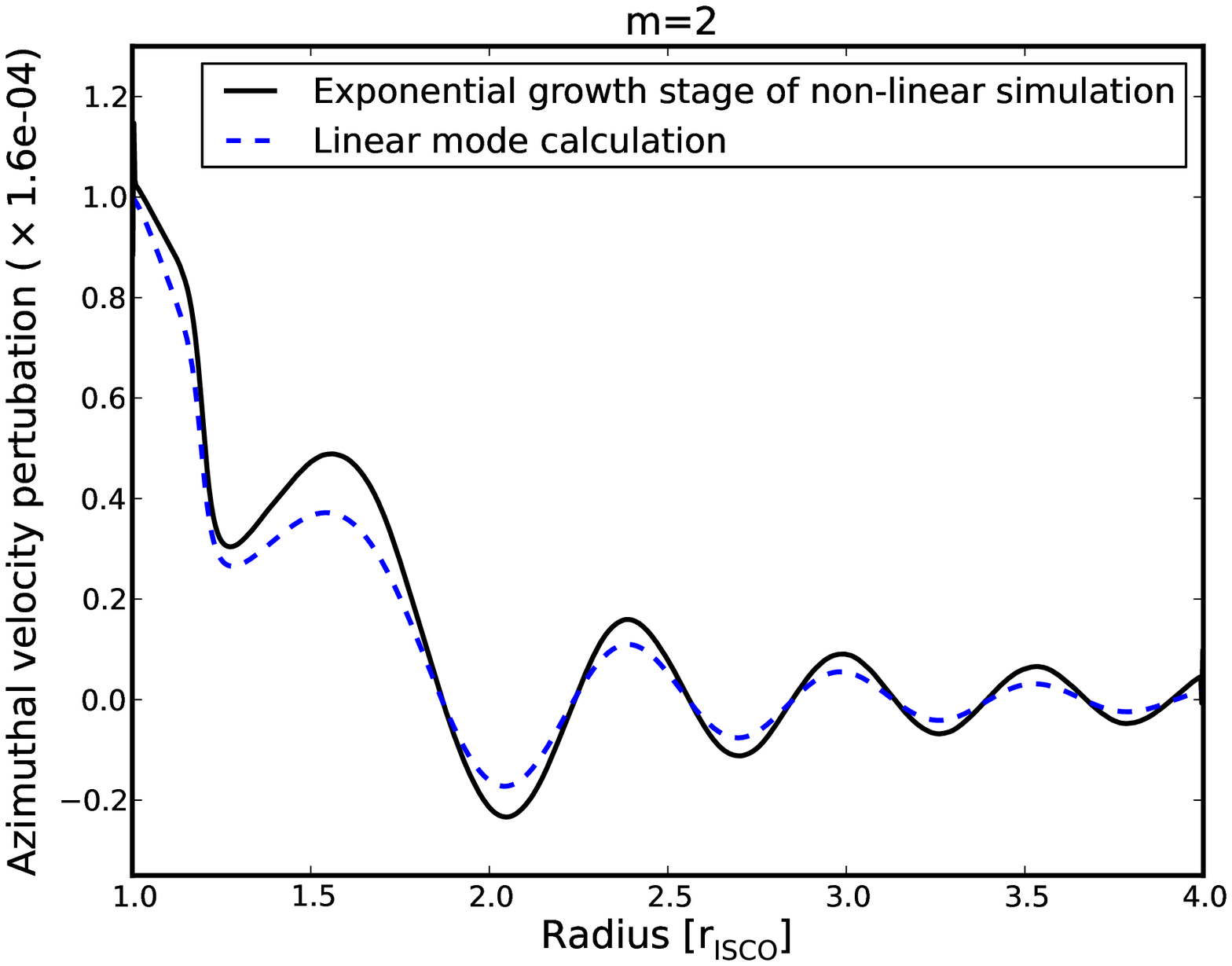} &
\includegraphics[width=0.45\textwidth]{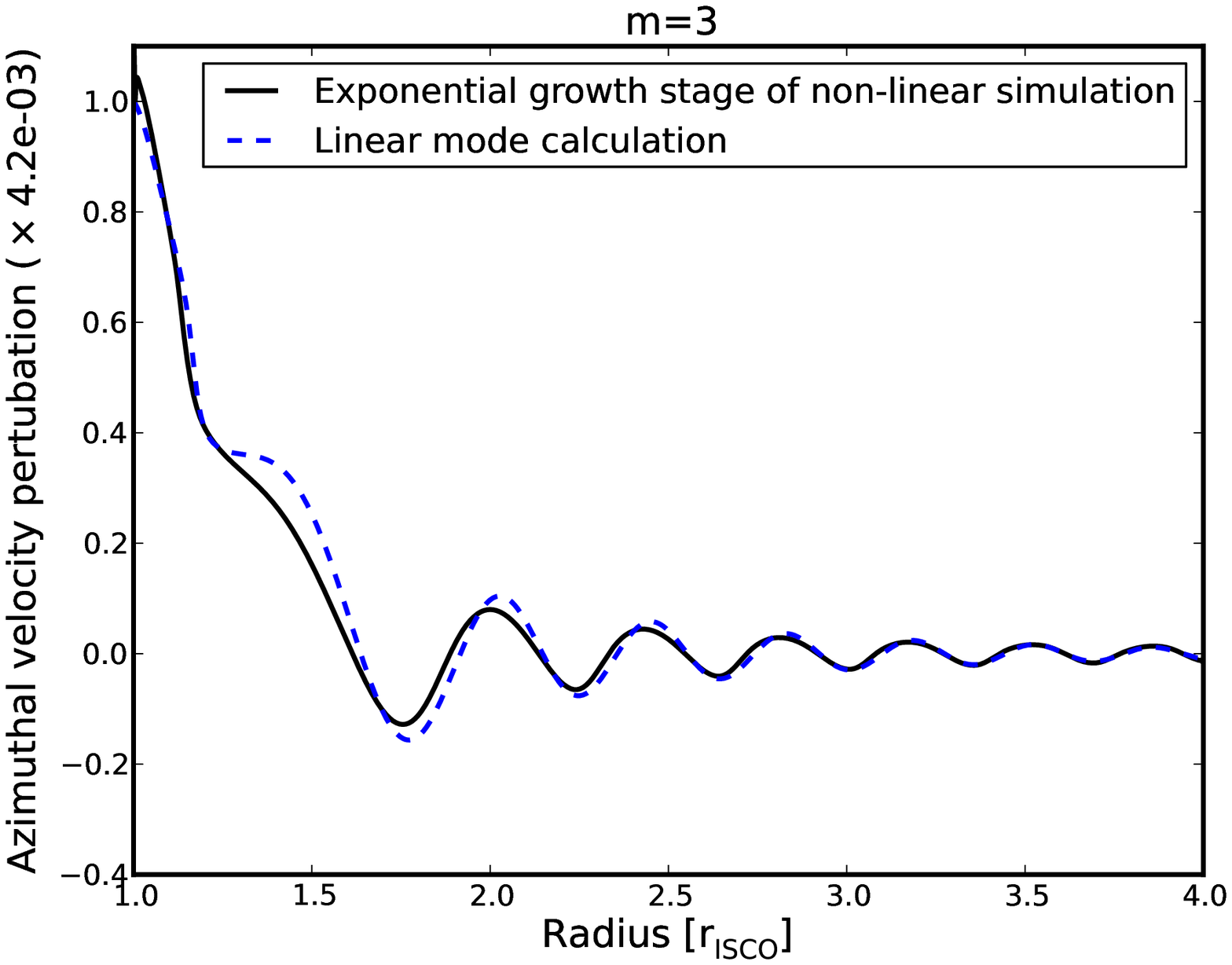} 
\end{array}
$
\caption{Comparison of the radial profiles of velocity perturbations
  from non-linear simulation and linear mode calculation. The top and
  bottom panels show the radial and azimuthal velocity perturbations,
  respectively.  The left and right panels are for cases with
  azimuthal mode number $m=2$ and $m=3$, respectively. In each panel,
  the dashed line is taken from the real part of the complex
  wavefunction obtained in linear mode calculation, while the solid line is from
  the non-linear simulation during the exponential growth stage (at
  $T=20$ orbits), with the quantities evaluated at $\phi=0.4\pi$.
  Note that the normalization factor is given in the $y$-axis label
  for the nonlinear simulation results.  }
\label{fig:vrvphi}
\end{center}
\end{figure*}

We solve the Euler equations that govern the dynamics of the disc flow
with the PLUTO code \footnote{publicly available at
  http://plutocode.ph.unito.it/} (Mignone et al. 2007), which is a
Godunov-type code with multiphysics and multialgorithm modules. For
this study, we choose a Runge-Kutta scheme (for time integration) and
piecewise linear reconstruction (for space integration) to achieve
second order accuracy, and Roe solver as the solution of Riemann
problems. The gird resolution we adopt is $(N_r\times
N_{\phi})=(1024\times 2048)$ so that each grid cell is almost
a square. Our runs typically last for 100 orbits (Keplerian orbits at
inner disc boundary). To compare with the linear mode calculations
(Lai \& Tsang 2009), the inner disk boundary is set to be reflective
with zero radial velocity. At the outer disc boundary, we adopt the
non-reflective boundary condition. This is realized by employing the
wave damping method (de Val-Borro et al. 2006) to reduce wave reflection. 
This implementation mimics the outgoing radiative boundary condition 
used in the linear analysis.

\section{Results}

We carried out simulations with two different types of initial
conditions for the surface density perturbation.  In the first, we
choose $\delta \Sigma(r, \phi)$ to be randomly distributed in $r$ and
$\phi$.  In the second, we impose $\delta \Sigma(r, \phi)\propto
\cos(m\phi)$ that is randomly distributed in $r$, so that the perturbation
has an azimuthal number $m$.
In all cases, the initial surface density perturbation has a small amplitude 
($|\delta\Sigma/\Sigma_{0}| \leq 10^{-4}$).

\begin{table*}
\caption{Comparison of results from linear and nonlinear studies of overstable disc p-modes}
\centering
\begin{threeparttable}
\begin{tabular}{ c c c c c c c c }
\noalign{\hrule height 1pt}
$m$\tnote{a} & $\omega_{r}$\tnote{b} &$\omega_i$\tnote{c} & $\omega_{r}/m\Omega$\tnote{d} & $\omega_{r1}$\tnote{e} & $|\omega_r-\omega_{r1}|/\omega_r$\tnote{f} & $\omega_{r2}$\tnote{g} & $|\omega_{r2}-\omega_{r1}|/\omega_{r2}$\tnote{h} \\
\hline
$2$ & $1.4066$ & $0.0632$ & $0.7033$ & $1.3998$ & $0.5\%$ & $1.4296$ & $2.1\%$ \\
$3$ & $2.1942$ & $0.0733$ & $0.7314$ & $2.1997$ & $0.3\%$ & $2.2445$ & $2.0\%$ \\
$4$ & $3.0051$ & $0.0763$ & $0.7512$ & $2.9996$ & $0.2\%$ & $3.0594$ & $2.0\%$ \\
$5$ & $3.8294$ & $0.0751$ & $0.7659$ & $3.7995$ & $0.8\%$ & $3.8886$ & $2.3\%$ \\
$6$ & $4.6621$ & $0.0714$ & $0.7770$ & $4.6494$ & $0.3\%$ & $4.7749$ & $2.6\%$\\
$7$ & $5.5007$ & $0.0664$ & $0.7858$ & $5.4992$ & $0.03\%$ & $5.6756$ & $3.1\%$ \\
$8$ & $6.3436$ & $0.0607$ & $0.7930$ & $6.3492$ & $0.09\%$ & $6.3189$ & $0.5\%$\\
\noalign{\hrule height 1.2pt}
\end{tabular}
\begin{tablenotes}
\item[a] Azimuthal mode number
\item[b] Mode frequency from the linear calculation (in units of Keplerian
  orbital frequency at the inner disc boundary; same for $\omega_i$,
  $\omega_{r1}$ and $\omega_{r2}$)
\item[c] Mode growth rate from the linear calculation 
\item[d] Ratio of wave pattern speed to the Keplerian orbital frequency at the inner disc boundary
\item[e] Mode frequency during the exponential growth stage of nonlinear simulation 
(peak frequency of the power density spectrum between $\sim 10$ orbits and $\sim 30$ orbits)
\item[f] Difference between $\omega_r$ (linear result) and $\omega_{r1}$ (nonlinear result)
\item[g] Mode frequency during the saturation stage of nonlinear simulation 
(peak frequency of the power density spectrum between $\sim 30$ orbits and $\sim 100$ orbits)
\item[h] Difference between $\omega_{r1}$ and $\omega_{r2}$
\end{tablenotes}
\end{threeparttable}
\label{tab:tab1}
\end{table*}

Fig.~\ref{fig:growth} shows the evolution of the radial velocity
amplitude near the inner disc radius for runs with initial azimuthal
mode number $m=2$, $m=3$ and random initial perturbation,
respectively. This velocity amplitude is obtained by searching for the
maximum $|u_r|$ at a given $r$ by varying $\phi$ for each given time
point. We chose $r=1.1$ because this is where the largest radial
velocity perturbation is located (see the upper panels of
Fig.~\ref{fig:vrvphi}).  We see that in all cases there are three
stages in the amplitude evolution.  The first stage occupies roughly
the first $10$ orbits, during which the initial perturbation starts to
affect the flow and presumably excites many modes/oscillations in the
disc. In the second stage (from $\sim 10$ orbits to $\sim
30$ orbits), the fastest growing mode becomes dominant and undergoes 
exponential growth with its amplitude increased by about four orders
of magnitude. In the last stage (beyond $\sim 30$ orbits), the
perturbation growth saturates and its amplitude remains at
approximately the same level.  A fit to the exponential growth stage
gives the growth rate of the fastest growing perturbation, which is
the slope of the fitted straight line. For the $m=2$ run, we find 
$0.17/\log_{10}(e)/2\pi\simeq 0.0637$ (in units of the orbital
frequency at $r=1$) as the growth rate, which is quite consistent with
the result from our linear eigenmode calculation $\omega_{i}\simeq
0.0632$ (the imaginary part of the eigenfrequency; Lai \& Tsang 2009;
Fu \& Lai 2011; see Table~\ref{tab:tab1}).  For the $m=3$ run,
our simulation gives $0.074$ as the mode growth rate, close to
$\omega_{i}=0.0733$ from the linear calculation. In
Fig.~\ref{fig:vrvphi}, we plot the radial profile of the velocity
perturbation at a given time during the exponential growth stage
of the simulation, and we compare it with the wavefunctions obtained
from the linear mode calculation. Note that in each panel of
Fig.~\ref{fig:vrvphi} we have normalized the different sets of data so
that they have the same scale. The normalization factor has also been
included in figure labels for easy recovery of the absolute value
in the simuation.  We can see that wavefunctions obtained from two studies 
agree quite well. The only obvious differences are near the outer disc boundary,
which can be attributed to the fact that outer boundary conditions
employed in the numerical simulation and linear mode calculation are not
exactly the same. Nevertheless, this agreement and the agreement in 
the mode growth rate in Fig.~\ref{fig:growth} confirm that our non-linear simulations 
indeed capture the same unstable disc p-modes as those revealed in linear
perturbation analysis.

\begin{figure*}
\begin{center}
$
\begin{array}{cc}
\includegraphics[width=0.4\textwidth]{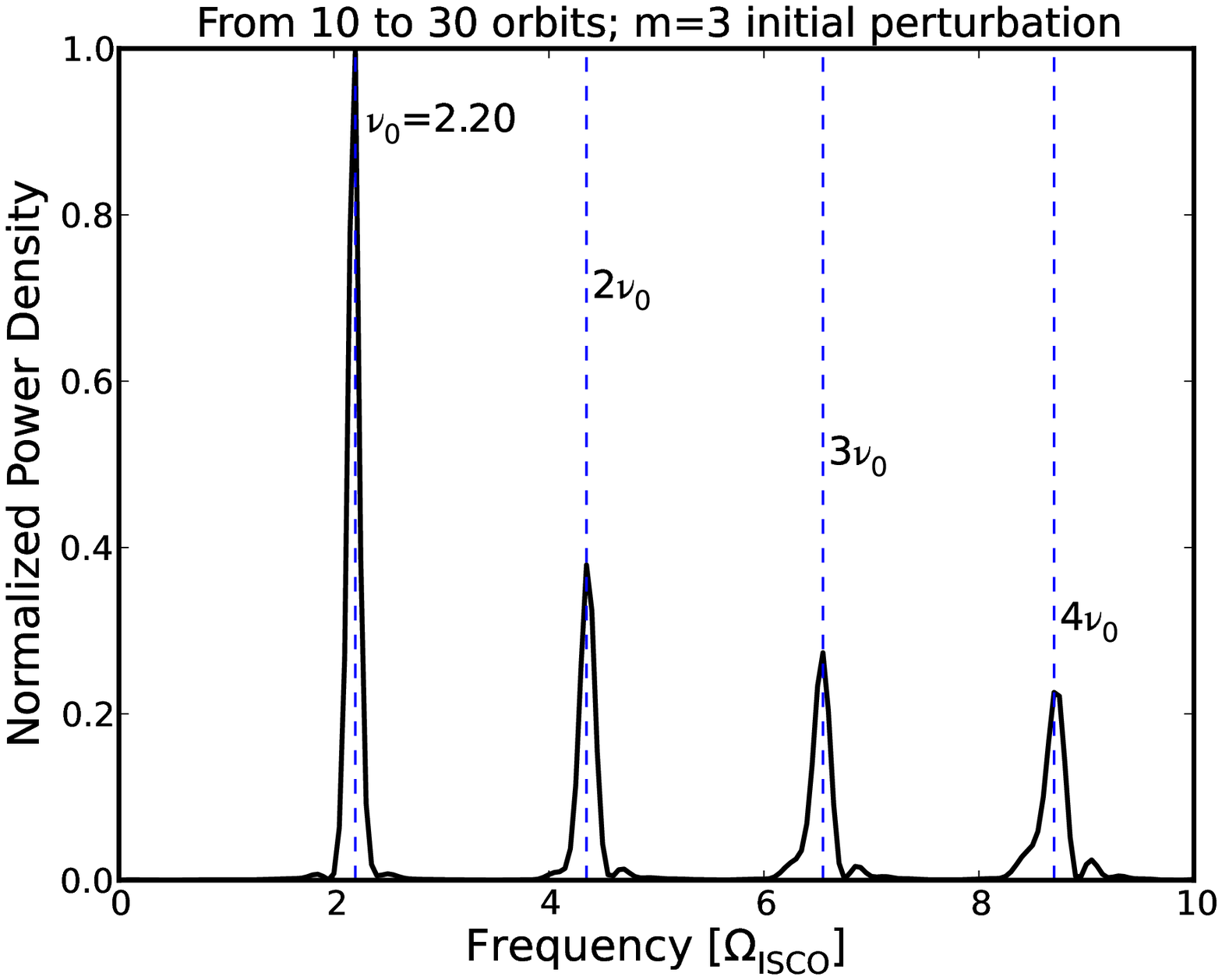} &
\includegraphics[width=0.4\textwidth]{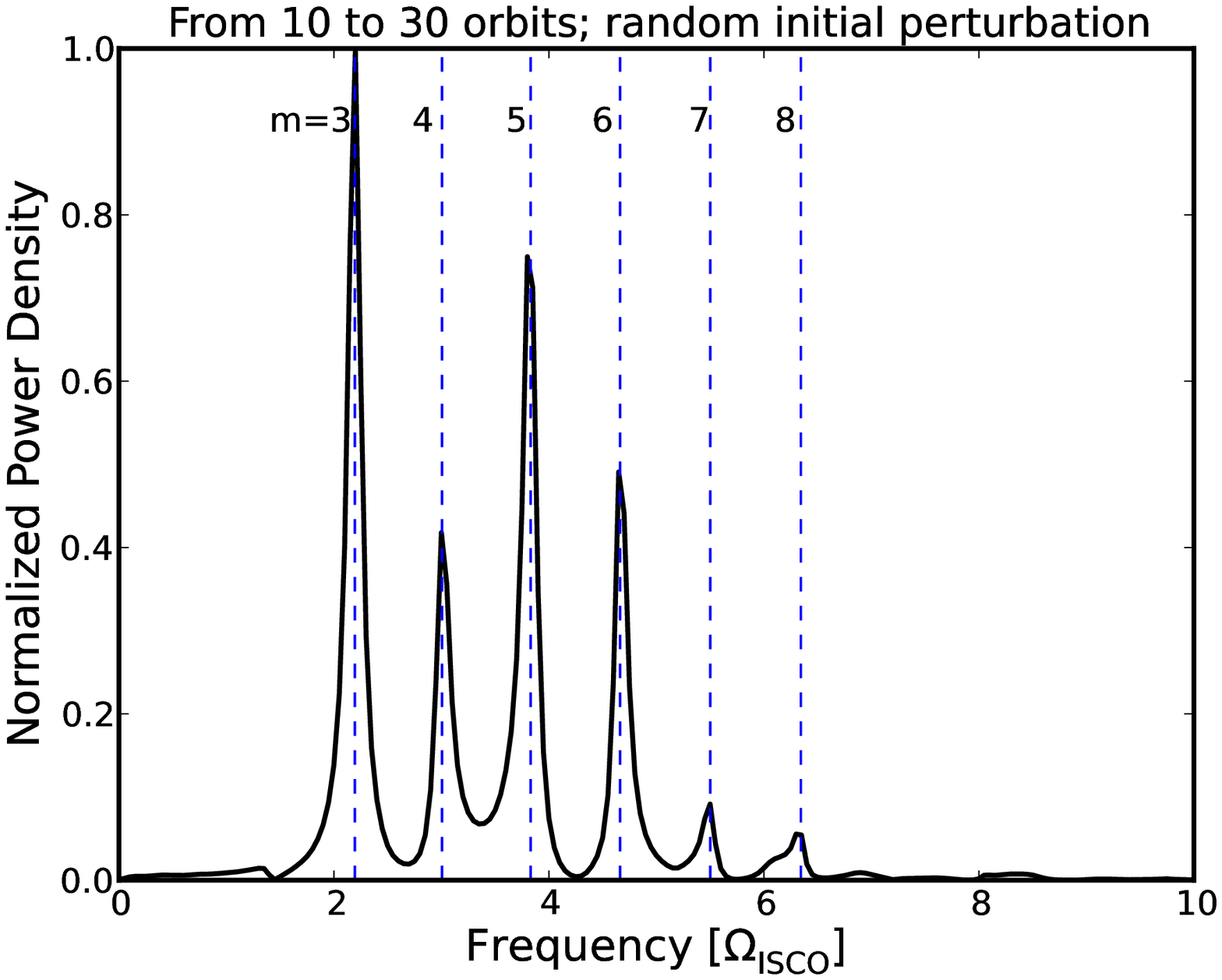} \\
\includegraphics[width=0.4\textwidth]{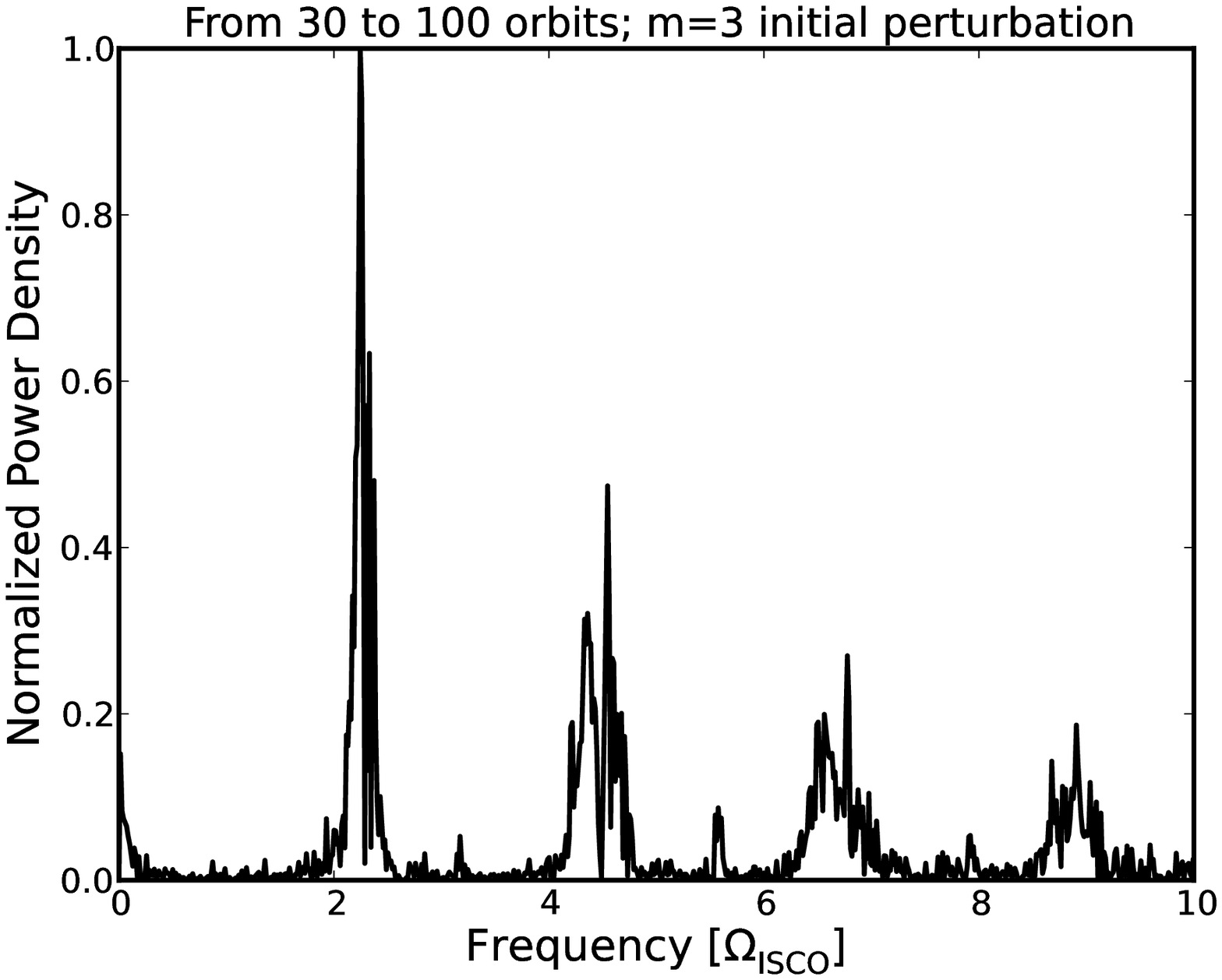} &
\includegraphics[width=0.4\textwidth]{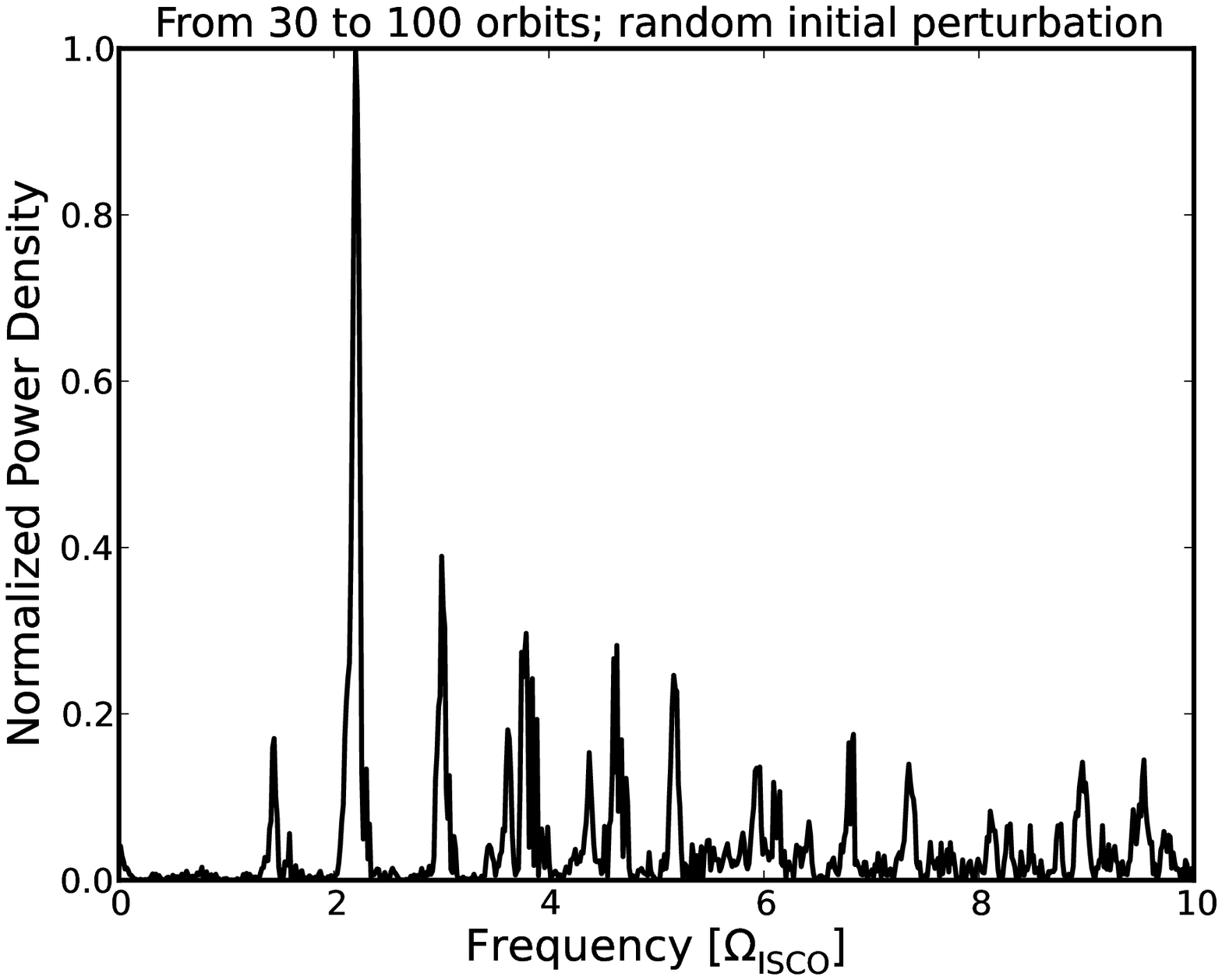} 
\end{array}
$
\caption{Power density spectra of the radial velocity perturbations
near the disk inner boundary ($r=1.1$). Each panel shows the normalized FFT
  magnitude as a function of frequency. The left and right
  columns are for runs with initial $m=3$ and with random perturbation,
  respectively. In the top and bottom panels, the Fourier transforms are
  sampled for time periods of [10, 30] orbits and [30,
    100] orbits, respectively.}
\label{fig:fft}
\end{center}
\end{figure*}

\begin{figure*}
\begin{center}
$
\begin{array}{cc}
\includegraphics[width=0.4\textwidth]{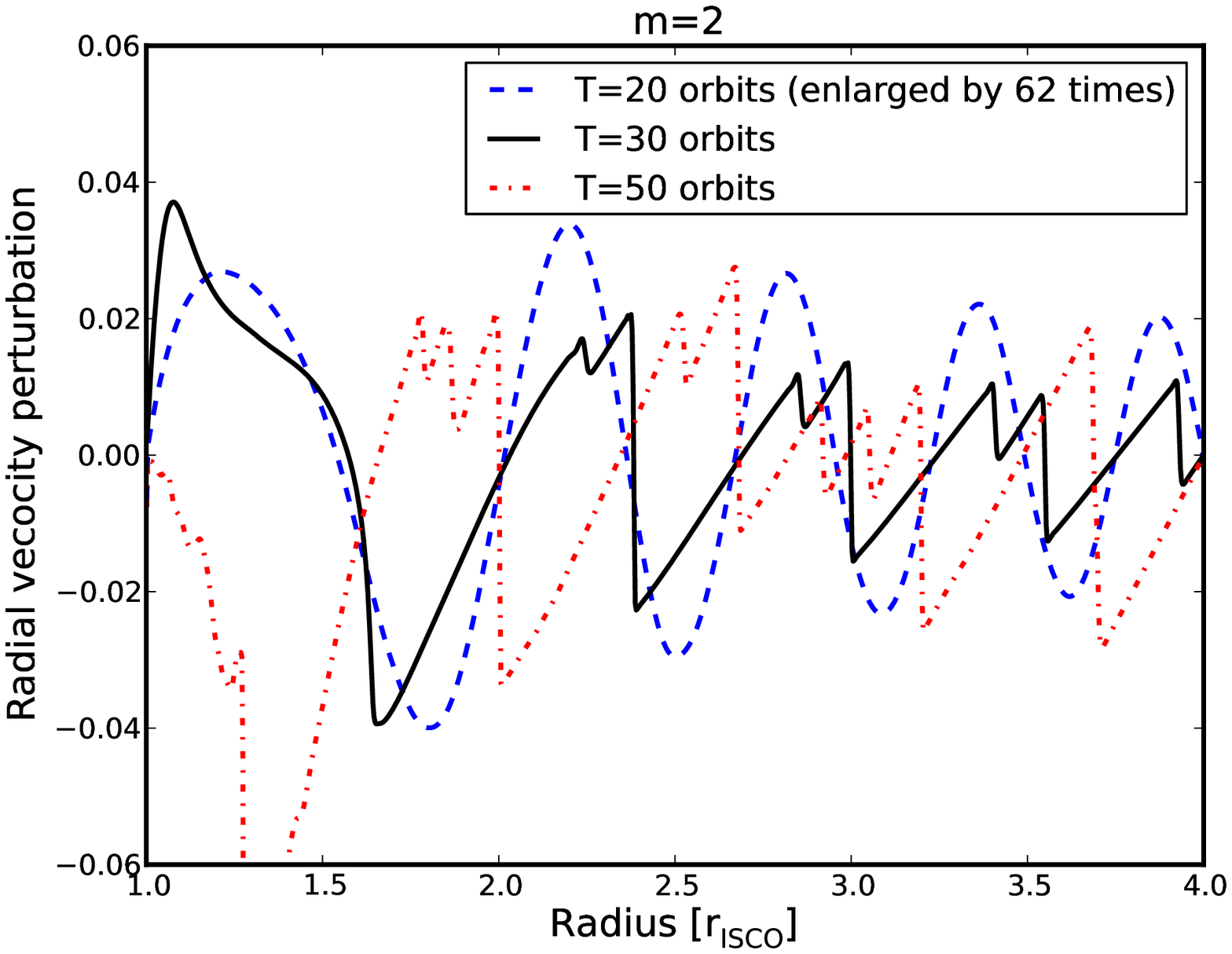} & 
\includegraphics[width=0.4\textwidth]{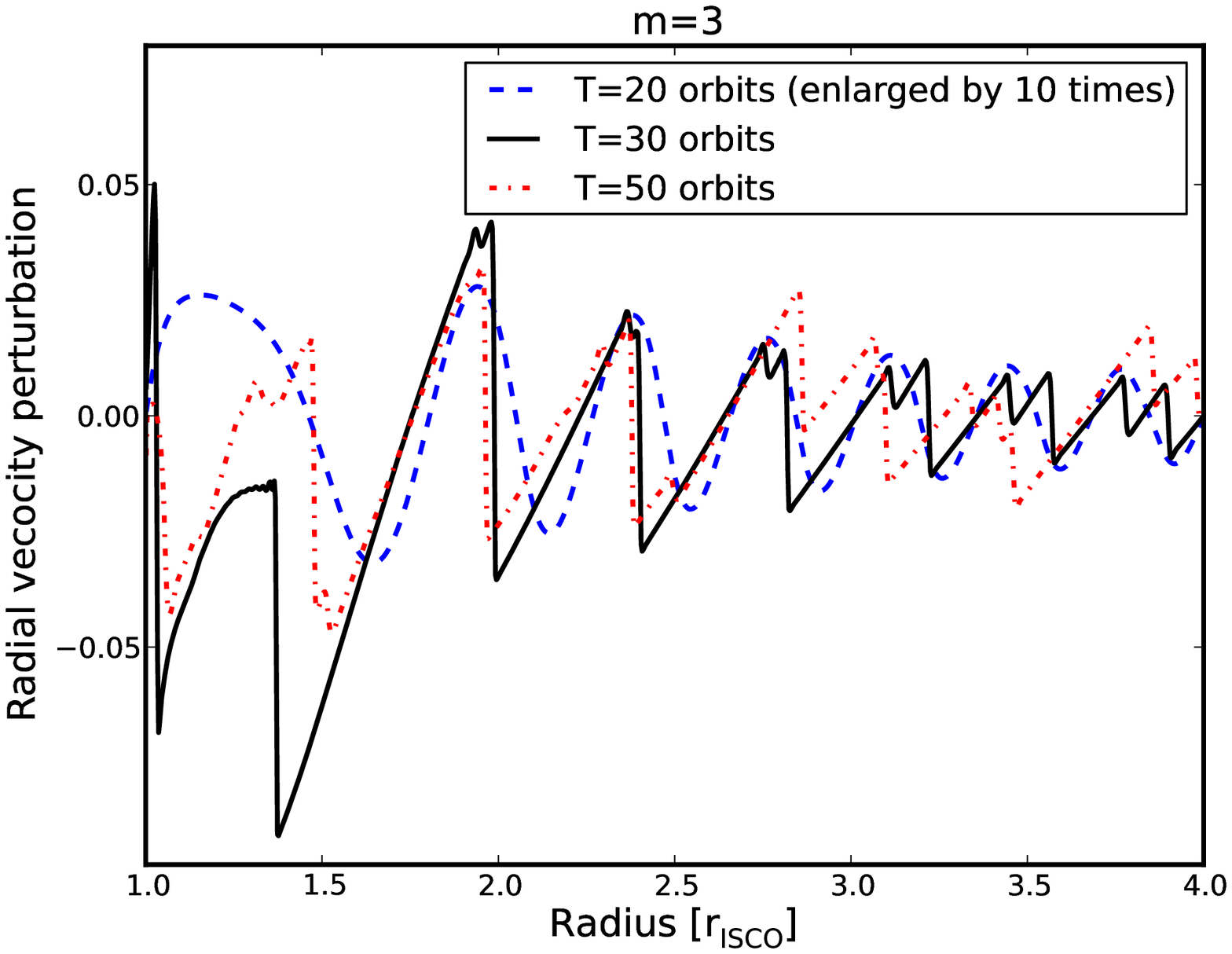} \\
\includegraphics[width=0.4\textwidth]{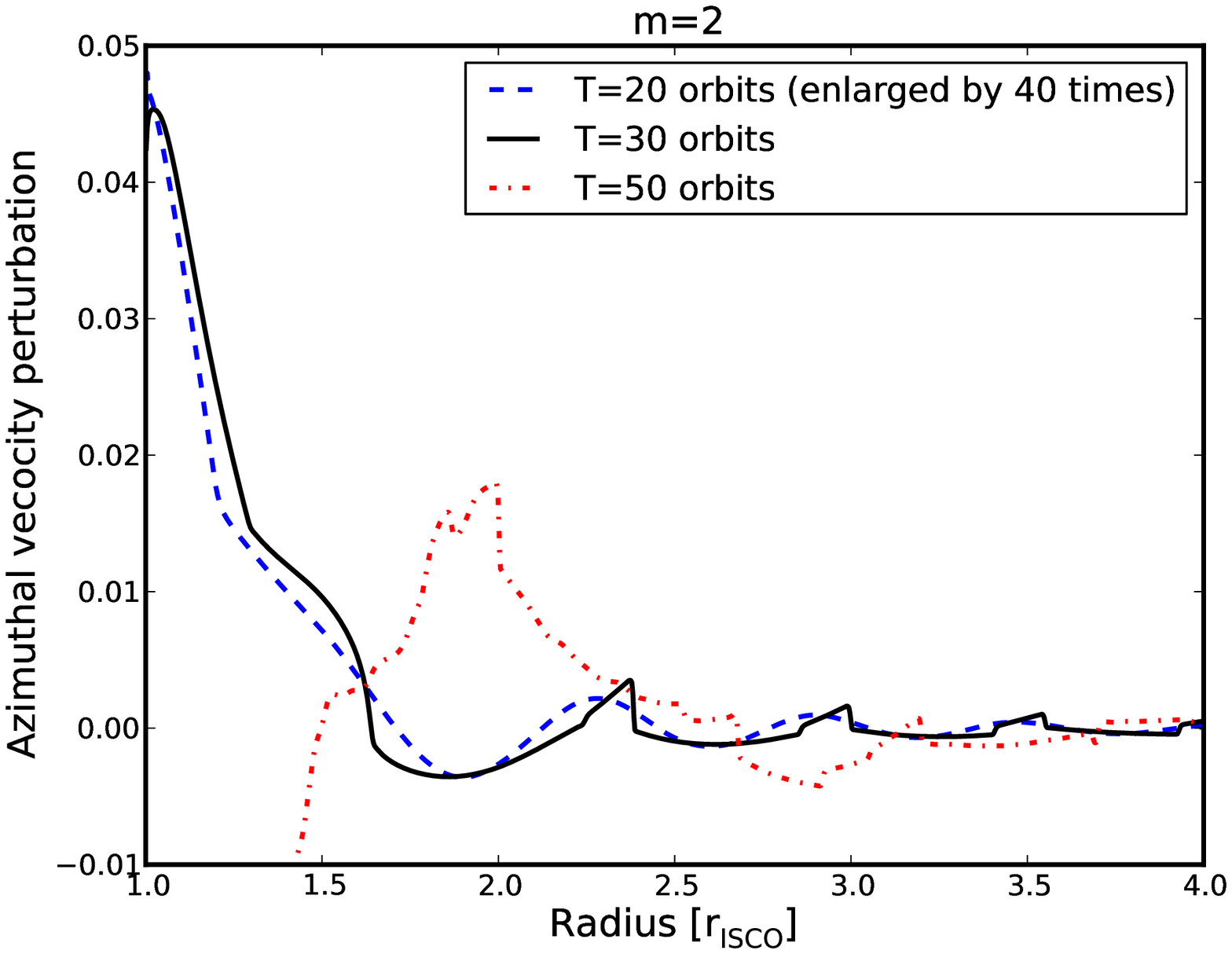} &
\includegraphics[width=0.4\textwidth]{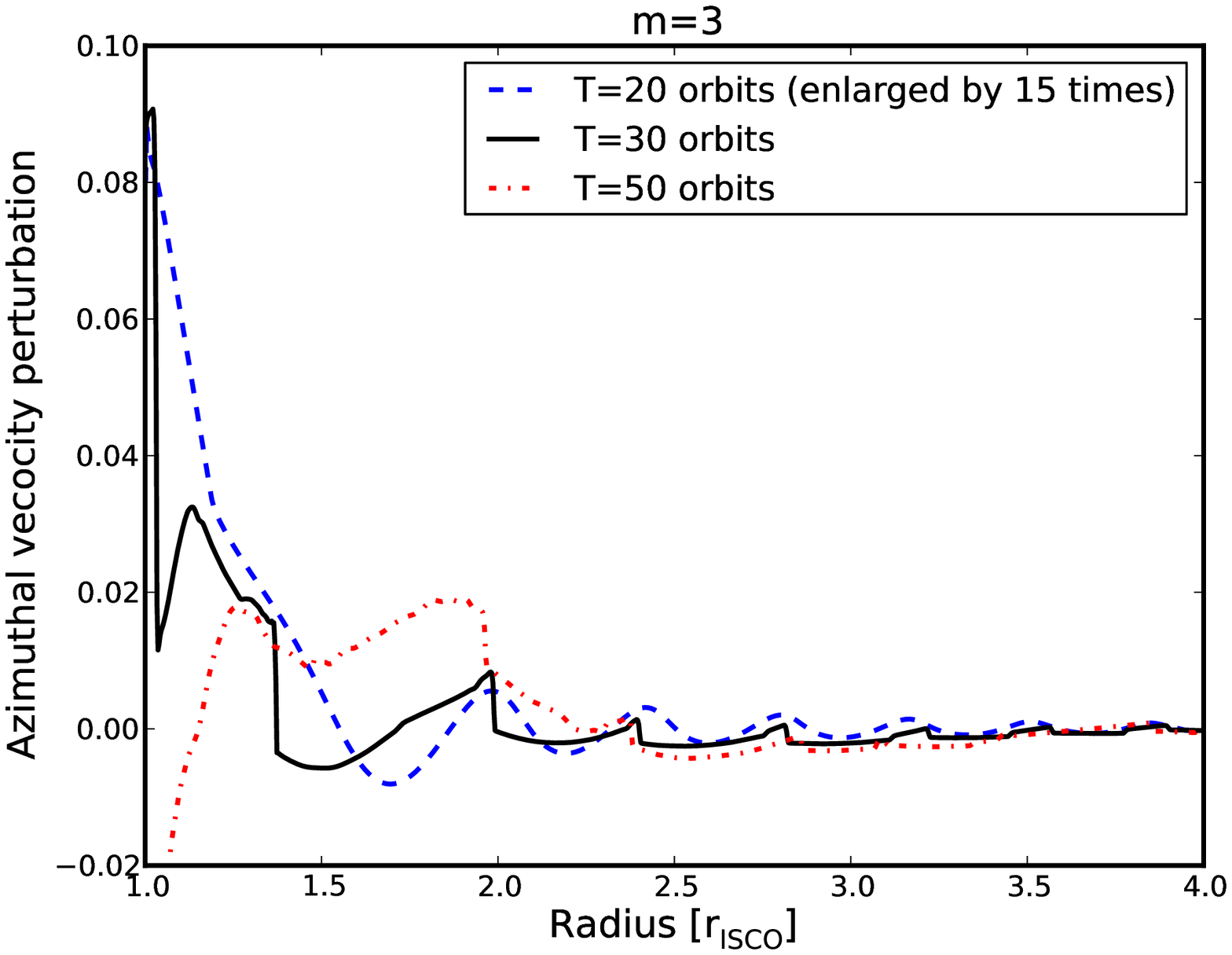} 
\end{array}
$
\caption{Evolution of the radial profiles of velocity perturbations from
  the non-linear simulations. The top and bottom panels show the radial
  and azimuthal components of velocity perturbation, respectively. 
  The left and right panels are for cases with azimuthal mode number
  $m=2$ and $m=3$, respectively. In each panel, the data are taken
  from points with fixed $\phi=0.4\pi$. Different line types represent
  different times during the simulation.}
\label{fig:cvrvphi}
\end{center}
\end{figure*}

\begin{figure*}
\begin{center}
$
\begin{array}{ccc}
\includegraphics[width=0.33\textwidth]{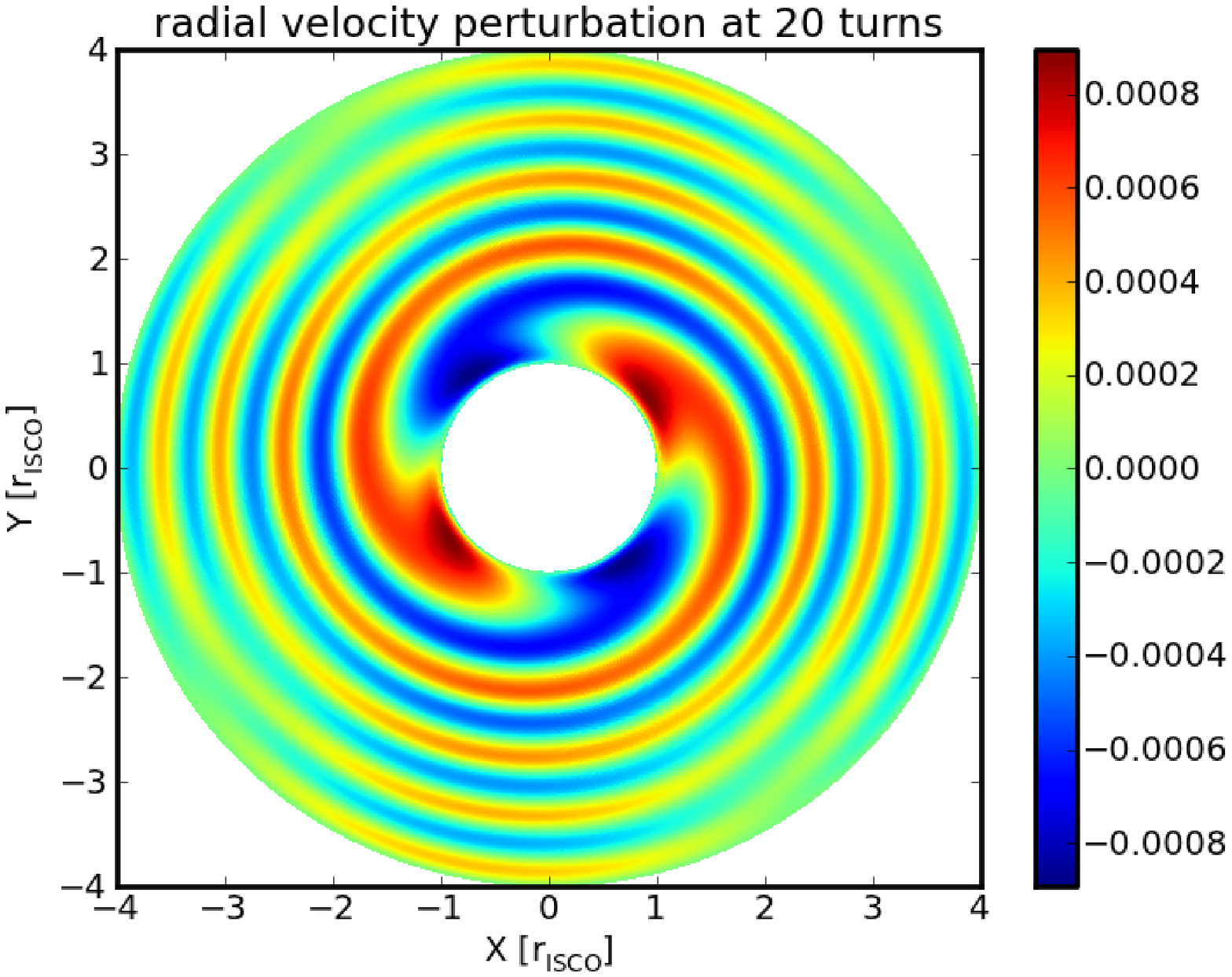} &
\includegraphics[width=0.33\textwidth]{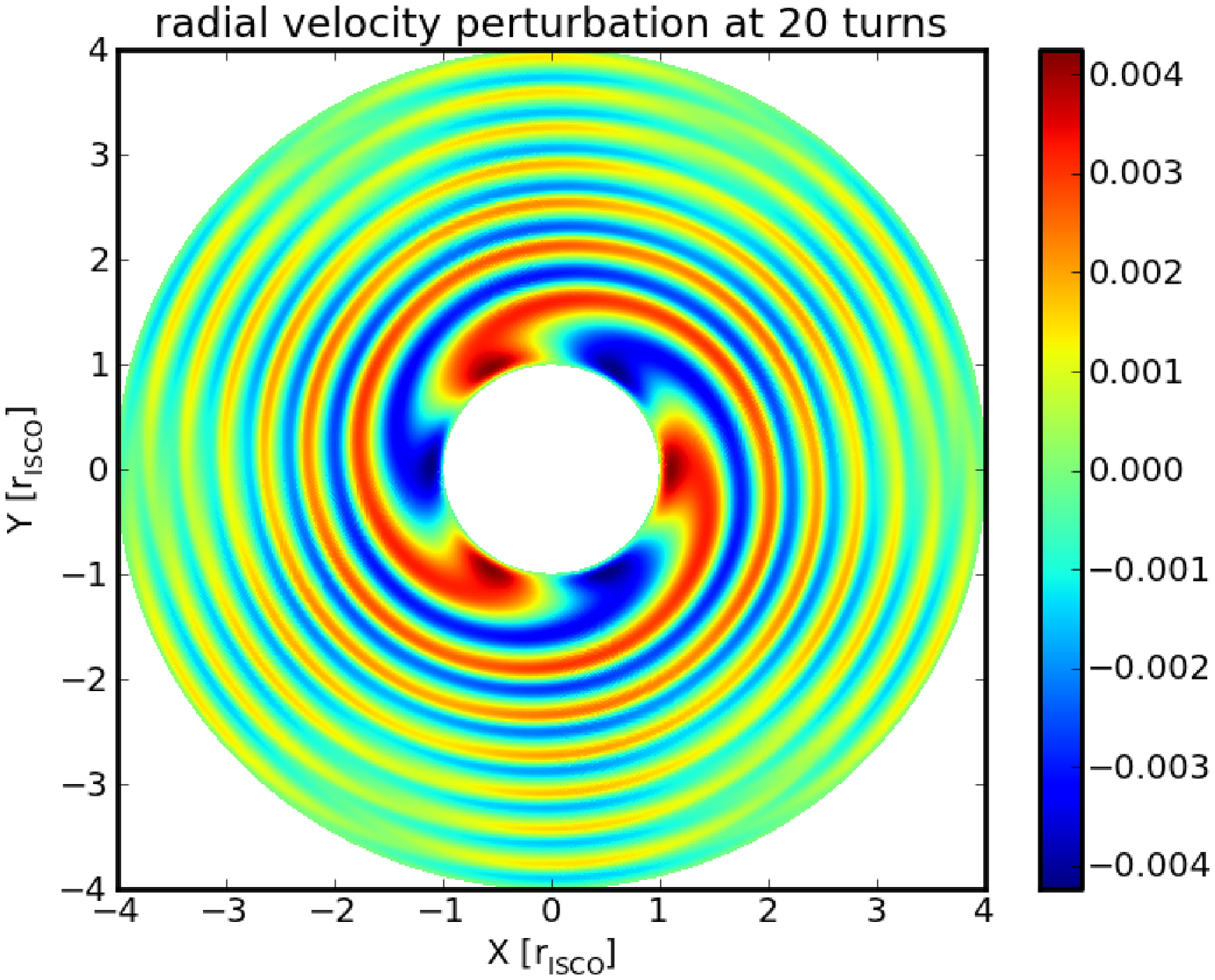} &
\includegraphics[width=0.33\textwidth]{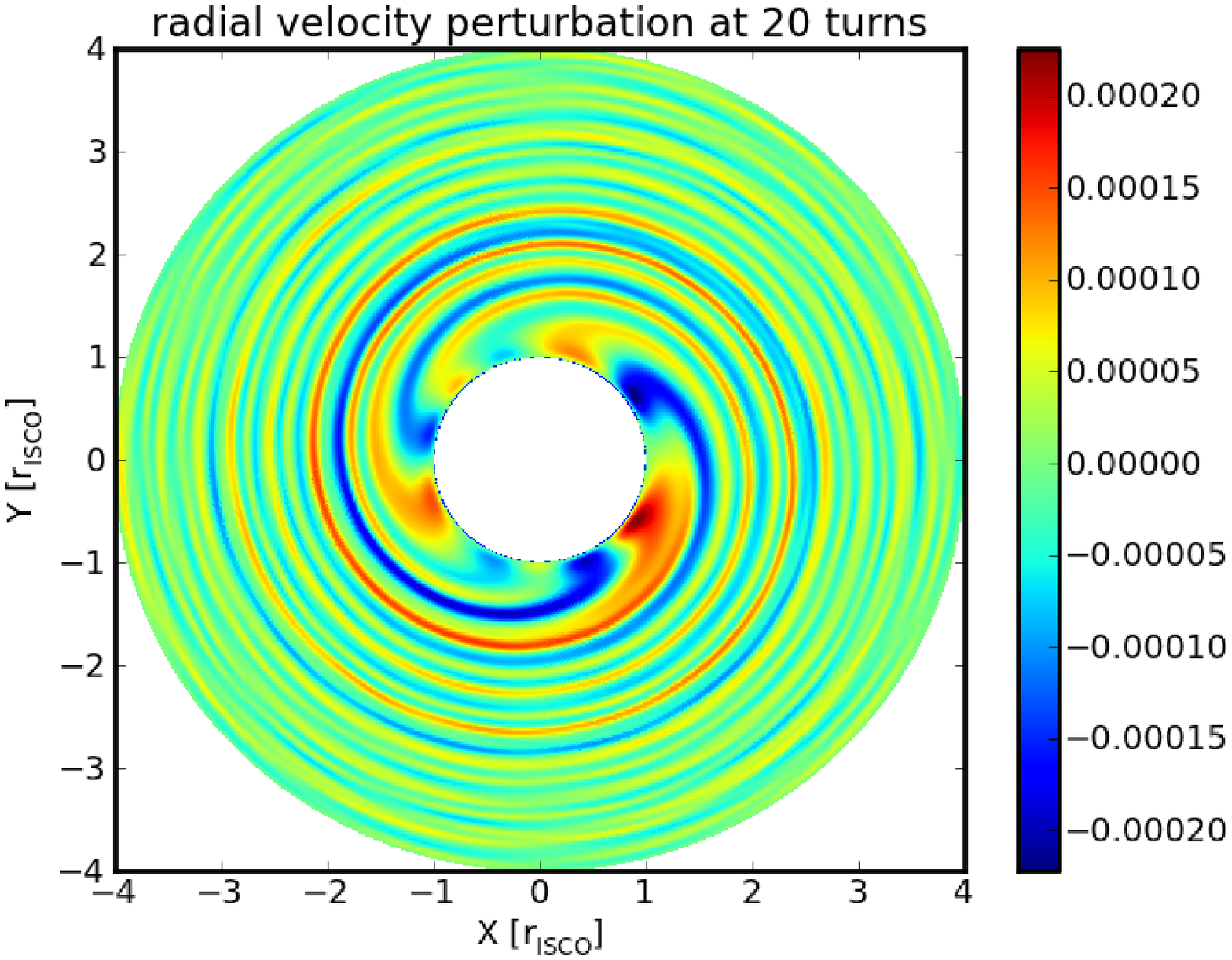} \\
\includegraphics[width=0.33\textwidth]{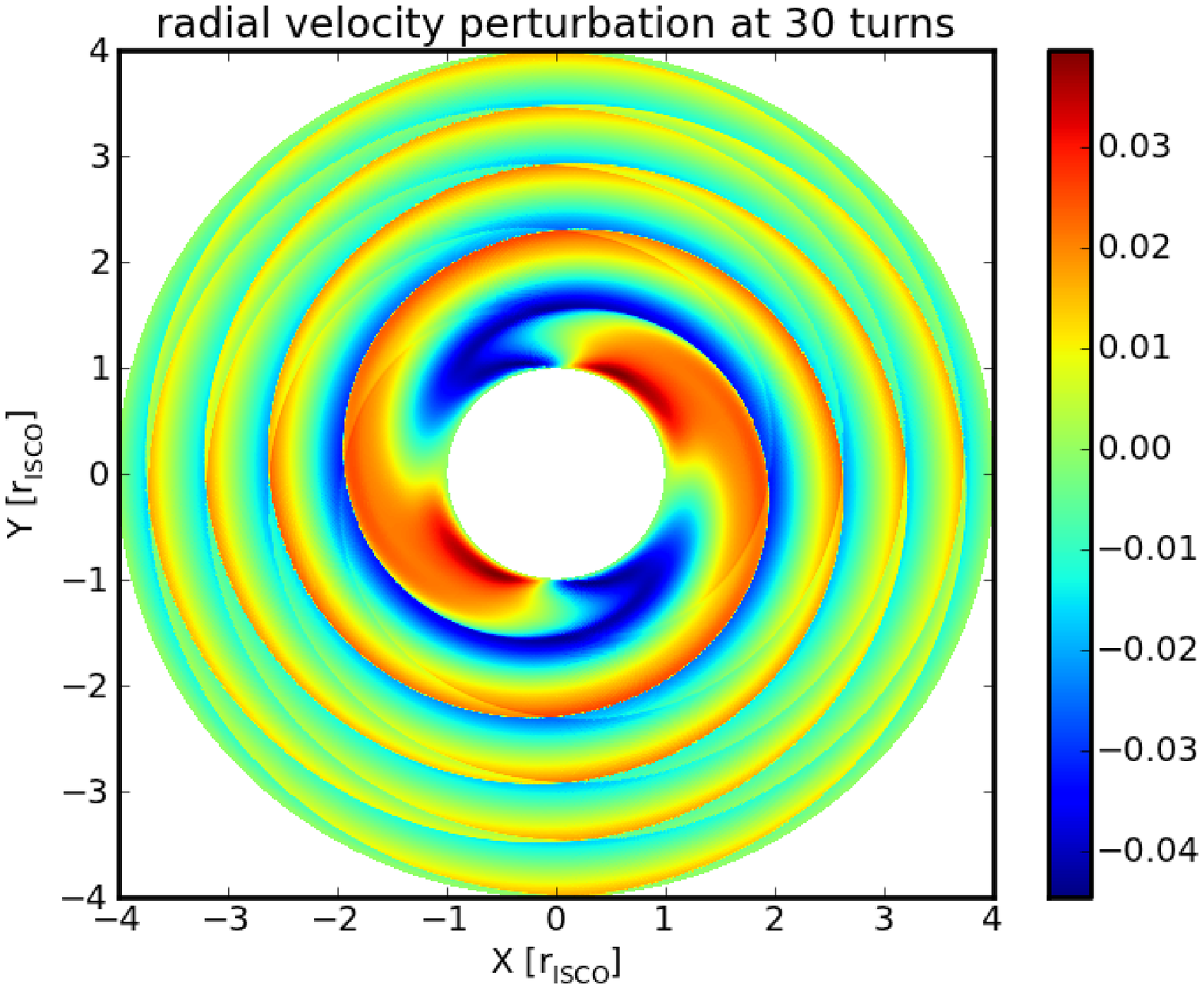} &
\includegraphics[width=0.33\textwidth]{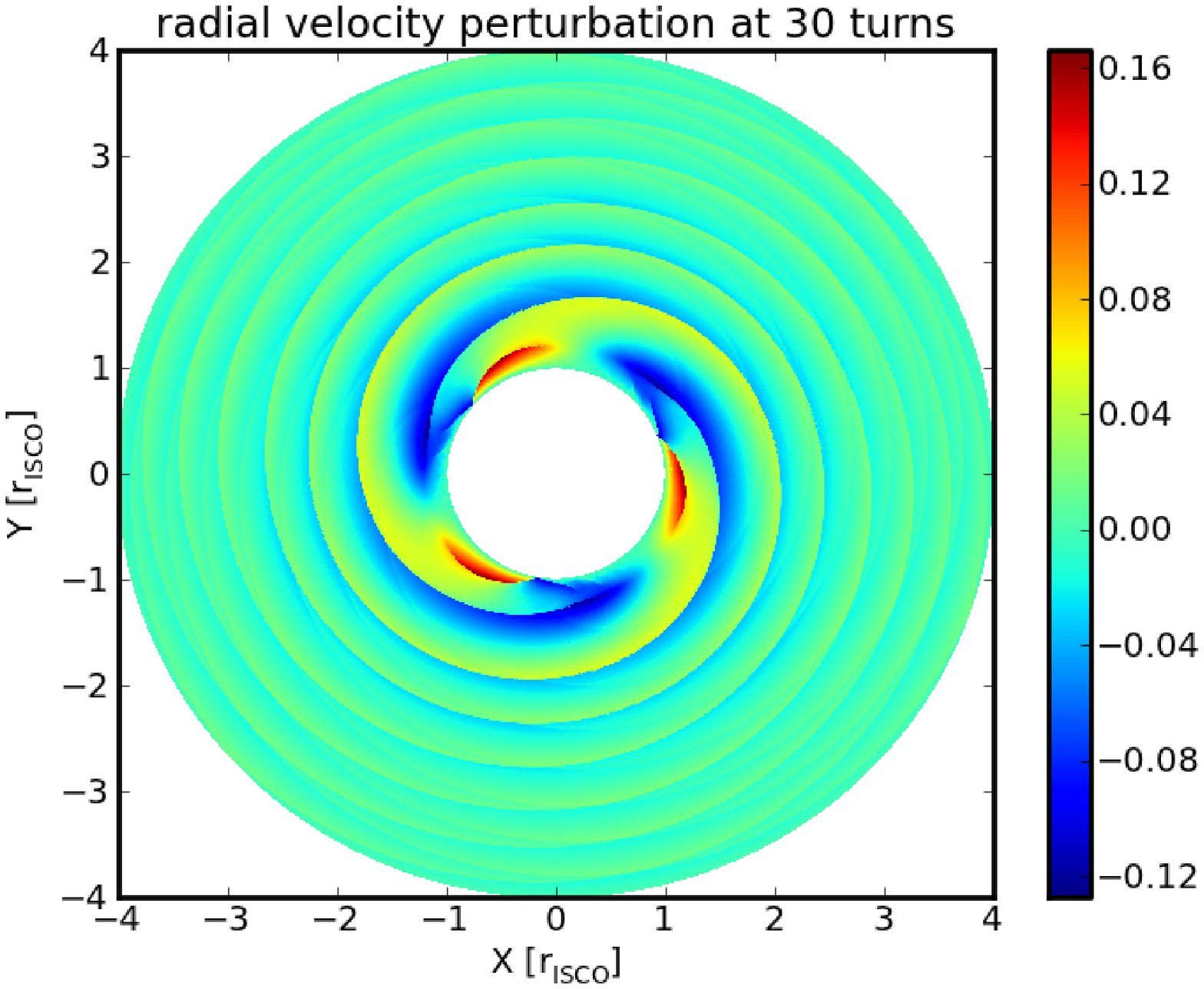} &
\includegraphics[width=0.33\textwidth]{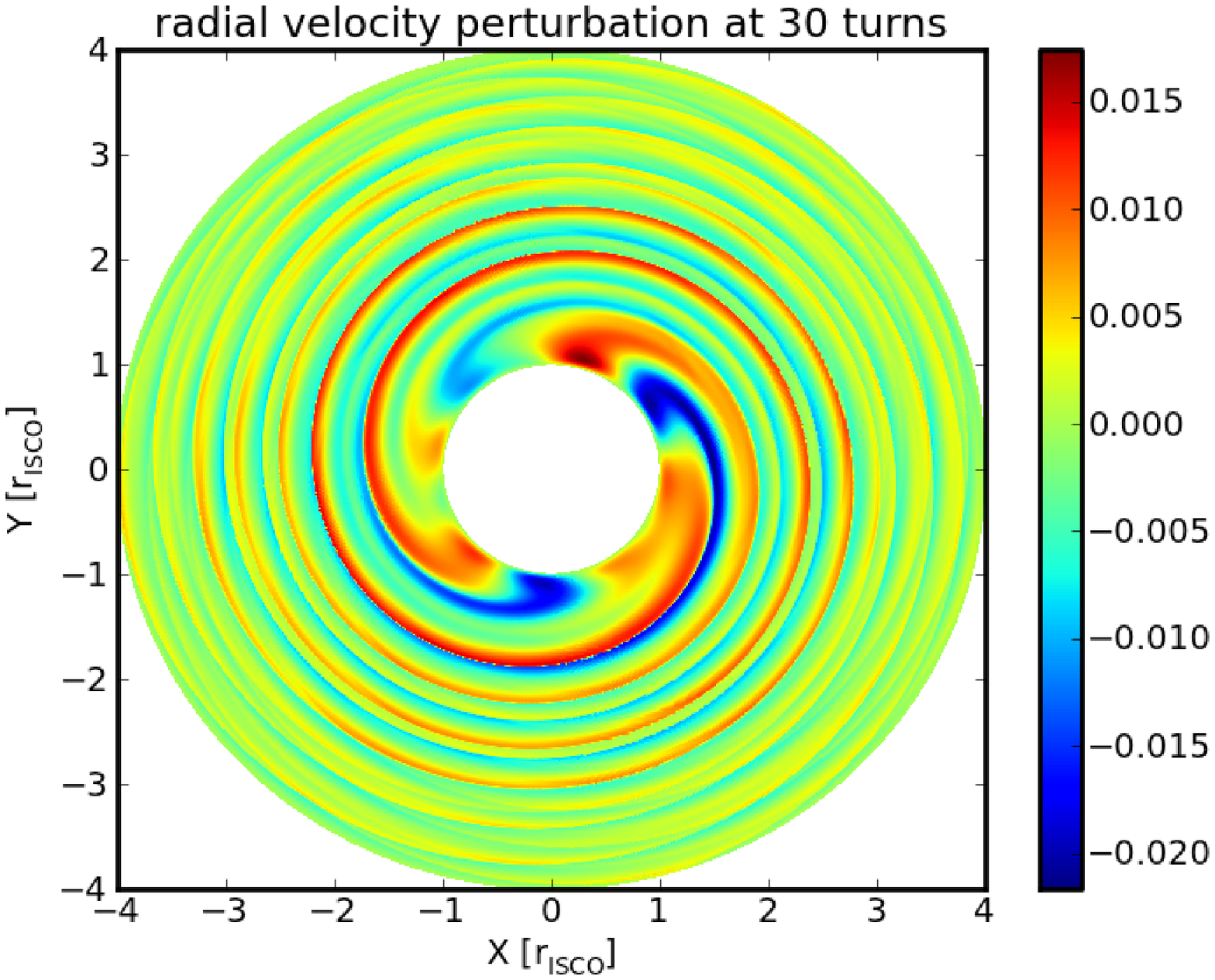} \\
\includegraphics[width=0.33\textwidth]{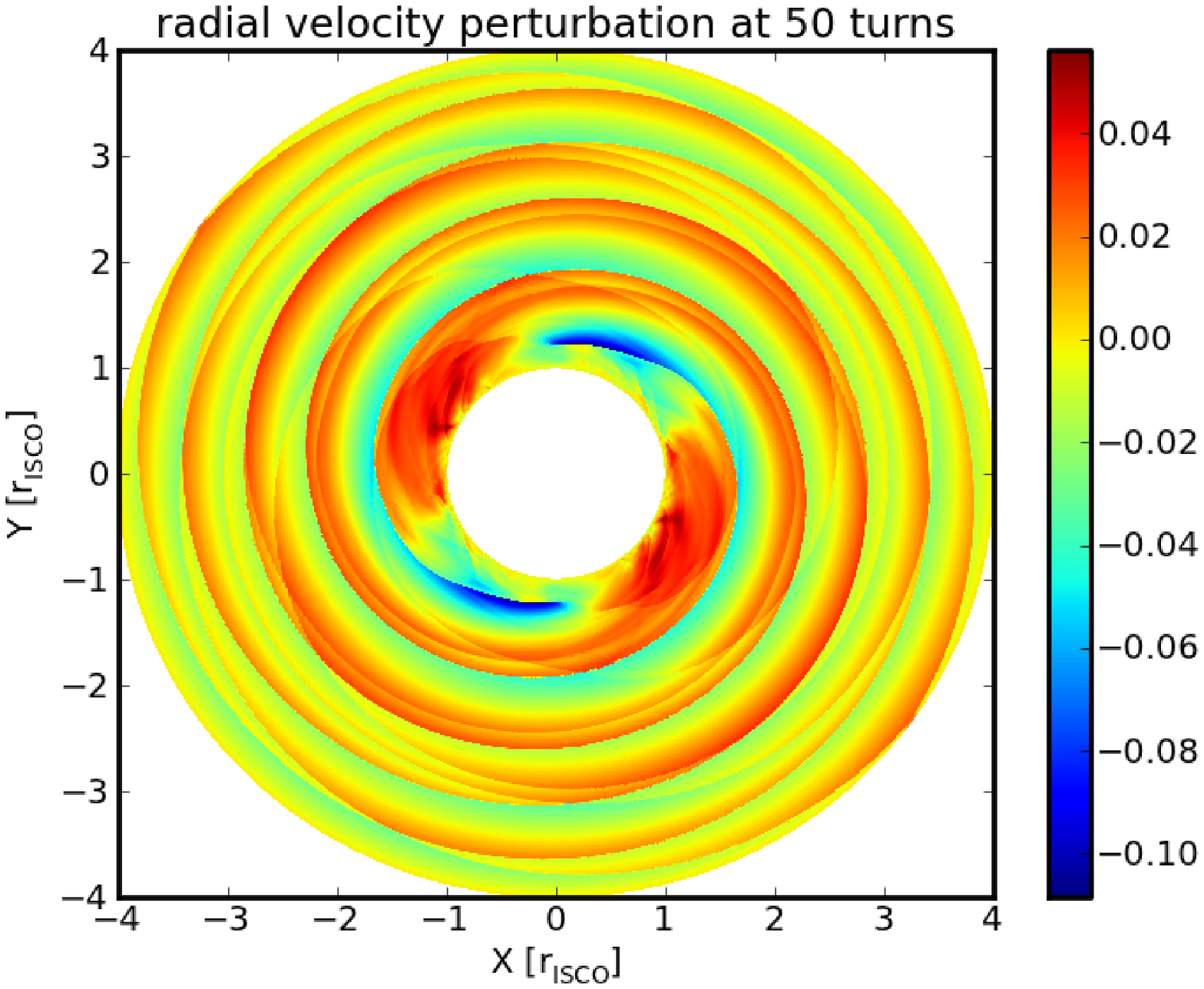} &
\includegraphics[width=0.33\textwidth]{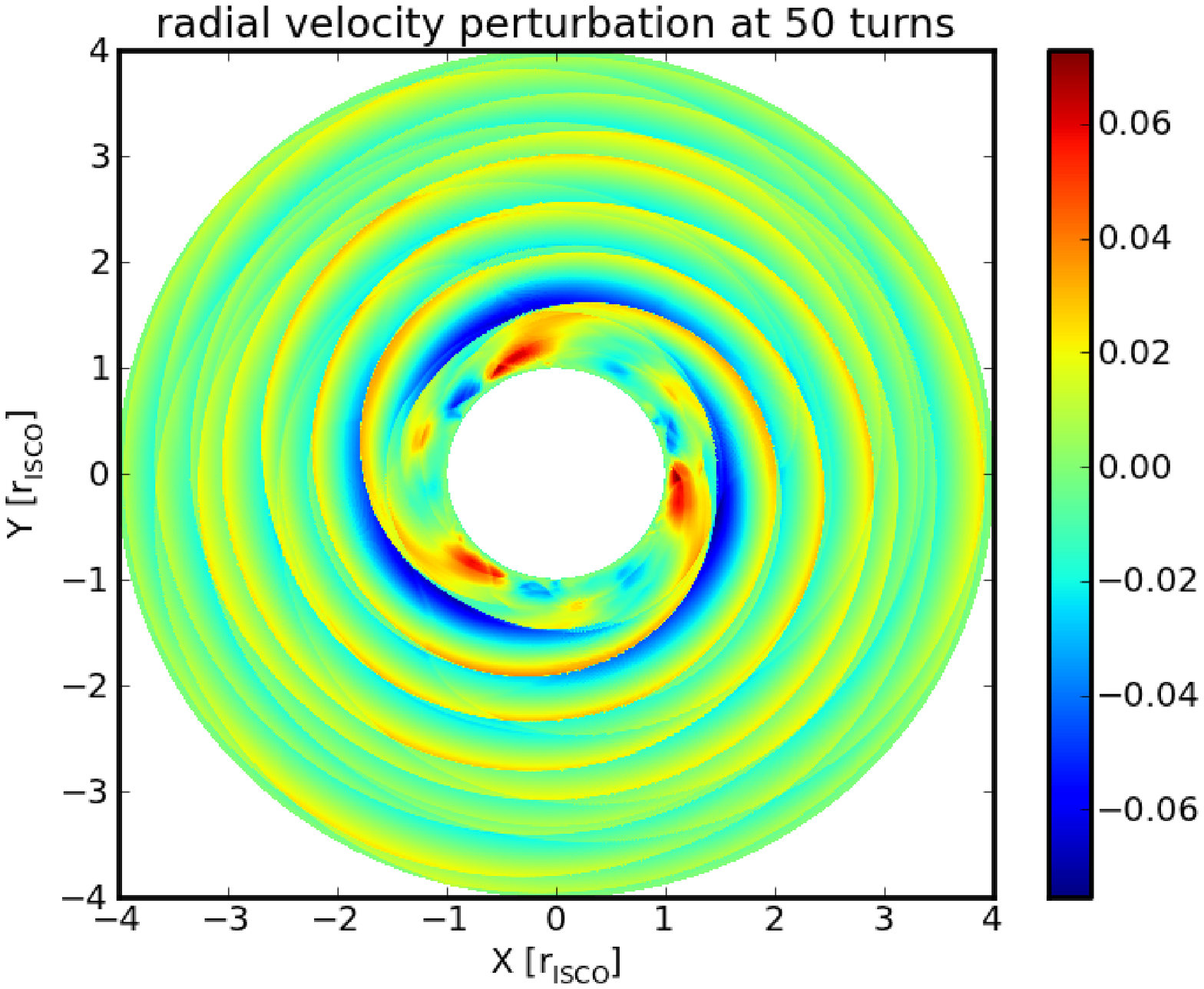}  &
\includegraphics[width=0.33\textwidth]{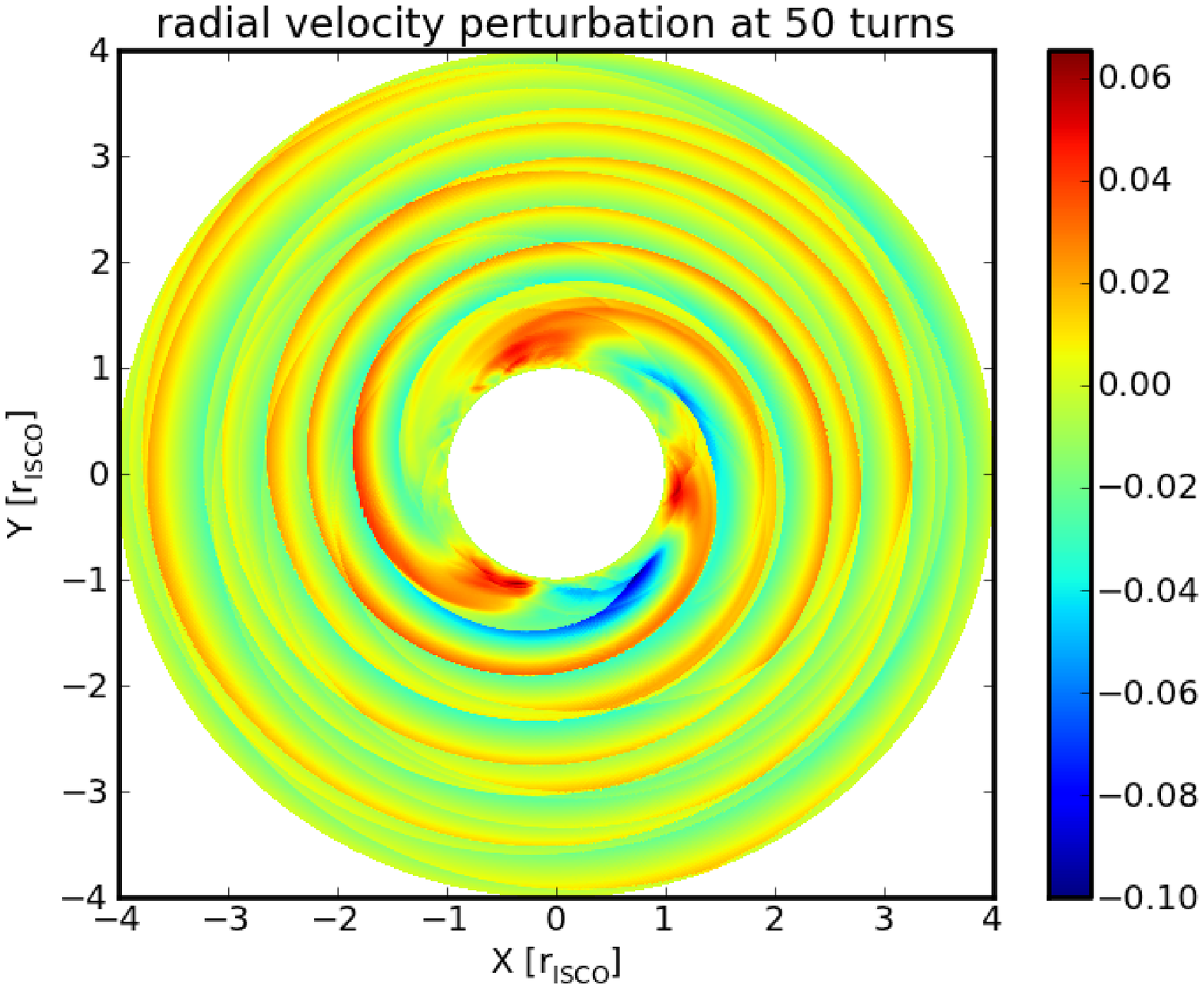} 
\end{array}
$
\caption{Evolution of the radial velocity for runs with initial $m=2$
  (left), $m=3$ (middle) and random (right) perturbations,
  respectively. From top to bottom, the times are $T=20$, $30$ and $50$ orbits, 
  respectively. Note that the color scale varies from panel to panel.}
\label{fig:vrcolor}
\end{center}
\end{figure*}

To explore the time variability of the flow, we carry out
Fourier transform of the radial velocity $u_r(r, \phi, t)$ at fixed
$r$ and $\phi$ during different evolutionary stages.  In
Fig.~\ref{fig:fft} we show some examples of the resulting power
density spectra (normalized to the maximum value of unity).  Different
rows correspond to different total sampling times and different
columns represent runs with different initial surface density
perturbation.  In the left columns, the disk has an initial
perturbation with azimuthal mode number $m=3$. A mixture of various
modes/oscillations are excited in the flow. After about $10$ orbits of
evolution, one of them (the fastest growing mode) has its oscillation
amplitude grown by a large amount such that it dominates over other
modes. This corresponds to the primary spike in the top-left
panel. The other spikes in the same panel are harmonics of this
primary spike (see the labels of the dashed vertical
lines). Table~\ref{tab:tab1} shows that the frequency\footnote{All the
  frequencies in this paper are angular frequencies unless otherwise
  noted.} of this fastest growing mode ($\omega_{r1}$) differs
from the frequency obtained in linear mode calculation by only
$0.3\%$, which again demonstrates the consistency of these two studies. 
After the perturbation saturates (bottom-left panel), we see
that the basic structure of the power density spectrum does not change
much except that these spikes are not as ``clean'' as in the linear regime;
this is probably due to the interaction of different modes. Compared with the upper
panel, the location of the primary spike ($\omega_{r2}$ in
Table~\ref{tab:tab1}) is increased by $2.0\%$. In
Table~\ref{tab:tab1} we also include the results from both linear mode
calculation and numerical simulation for modes with other mode number
$m$. The comparison illustrates two main points: First, the
frequencies of the fastest growing modes during the exponentially
growth stage of numerical simulations are exceptionally close to the
linear calculation results (differ by less than $1\%$); second, the
frequencies of the fastest growing modes during the saturation stage
are only slightly higher (except for the $m=8$ mode which shows lower
frequency) than ones during the exponential growing stage. These indicate
that the mode frequencies obtained in linear mode calculation are
fairly robust and can be reliably applied in the interpretation of HFQPOs.

In the right columns of Fig.~\ref{fig:fft}, the simulation starts with
a random initial perturbation which excites modes with various
$m$'s. During the exponential growth stage, six modes stand out. By
examining the location of the corresponding spikes, we know that these
are the $m=3$, $4$, $5$, $6$, $7$, $8$ modes. Although the
$m=3$ mode seems to be the most prominent one in this figure, we note
that this is because for this particular run the initial random
perturbation happens to contain more $m=3$ wave components. If we were
to start the run with a different initial random perturbation, then
the relative strengths of those peaks would also be different (not
necessarily with $m=3$ being dominant).

Fig.~\ref{fig:cvrvphi} shows the comparison of the velocity perturbations 
during the linear stage ($T=20$ orbits), at the end of the linear
stage ($T=30$ orbits) and during the saturation stage ($T=50$ orbits)
for simulations with different initial perturbations. At $T=20$
orbits, the oscillation mainly comes from the single fastest growing
mode and has a smooth radial profile. At $T=30$ orbits, the perturbation
starts to saturate, and the oscillation now consists of many different modes,
and its radial profile exhibits sharp variations at several locations.
These sharp features remain after the saturation ($T=50$ orbits).

To see this evolution from a different perspective, in
Fig.~\ref{fig:vrcolor} we show the color contours of radial velocity
for runs with different initial perturbations (different columns) at
different times (different rows). We can clearly see that spiral waves
gradually develop due to the instability. As the system evolves, sharp
features in velcoities emerge. At the end (the bottom 
row) the spiral arms becomes more irregular, which is related
to the emergence and interaction of multiple modes after saturation.

The sharp velocity variations shown in
Figs.~\ref{fig:cvrvphi}-\ref{fig:vrcolor} suggest shock-like features.
Note that since our simulations adopt isothermal equation of state
(with $P/\Sigma=$constant), there is no entropy generation and shock
formation in the strict sense.  The radial velocity jump (see
Fig.~\ref{fig:cvrvphi}) is comparable but always smaller than the
sound speed. Nevertheless, these sharp features imply that
wave steepening plays an important role in the mode saturation. 

\section{Conclusions}

We have carried out high-resolution, two-dimensional hydrodynamical
simulations of overtstable inertial-acoustic modes (p-modes) in BH
accretion discs with various initial conditions.  The evolution of
disc p-modes exhibits two stages.  During the first (linear) stage, the
oscillation amplitude grows exponentially.  In the cases with a
specific azimuthal mode number ($m$), the mode frequency, growth rate
and wavefunctions agree well with those obtained in our previous
linear mode analysis.  In the cases with random initial perturbation,
the disc power-density spectrum exhibits several prominent
frequencies, consistent with those of the fastest growing linear modes
(with various $m$'s).  These comparisons with the linear theory
confirm the physics of corotational instability that drives disc
p-modes presented in our previous studies (Lai \& Tsang 2009; Tsang \&
Lai 2009c; Fu \& Lai 2011, 2012).  In the second stage, the mode growth
saturates and the disc oscillation amplitude remains at roughly a
constant level.  In general, we find that the primary disc oscillation
frequency (in the cases with specific initial $m$) is larger than the
linear mode frequency by less than $4\%$, indicating the robustness of
disc oscillation frequency in the non-linear regime. Based on the
sharp, shock-like features of fluid velocity profiles, we suggest that
the nonlinear saturation of disc oscillations is caused by wave
steepening and mode-mode interactions.

As noted in Section 1, our 2D hydrodynamical simulations presented in
this paper do not capture various complexities (e.g., magnetic field,
turbulence, radiation) associated with real BH accretion discs.
Nevertheless, they demonstrate that under appropriate conditions, disc
p-modes can grow to nonlinear amplitudes with well-defined frequencies
that are similar to the linear mode frequencies.  A number of issues
must be addressed before we can apply our theory to the interpretation
of HFQPOs.  First, magnetic fields may play an important role in the
disc oscillations.  Indeed, we have shown in previous linear
calculations (Fu \& Lai 2011,2012) that the growth of p-modes can be
significantly affected by disc toroidal magnetic fields. A strong,
large-scale poloidal field can also change the linear mode frequency
(Yu \& Lai 2013). Whether or not these remain true in the non-linear regime is
currently unclear. Second, understanding the nature of the inner disc boundary
is crucial. Our calculations rely on the assumption that the inner disc edge
is reflective to incoming spiral waves. In the standard disc model with zero-torque
inner boundary condition, the radial inflow velocity is not negligible near the ISCO, 
and the flow goes through a transonic point. While the steep density and velocity
gradients at the ISCO give rise to partial wave reflection (Lai \& Tsang 2009),
such radial inflow can lead to significant mode damping such that the net growth rates of
p-modes become negative. This may explain the absence of HFQPOs in the thermal
state of BH x-ray binaries. However, it is possible that the inner-most region of 
BH accretion discs accumulates significant magnetic flux and forms a magnetosphere.
The disc-the magnetosphere boundary will be highly reflective, leading to the 
growth of disc oscillations (Fu \& Lai 2012; see also Tsang \& Lai 2009b).
Finally, the effect of MRI-driven disc turbulence on the p-modes
requires further understanding. In particular, turbulent viscosity 
may lead to mode growth or damping, depending on the magnitude and the 
density-dependence of the viscosity (R. Miranda \& D. Lai 2013, in prep).

\section*{Acknowledgements}
This work has been supported in part by the NSF grants AST-1008245,
AST-1211061 and the NASA grant NNX12AF85G.  WF also acknowledges the
support from the Laboratory Directed Research and Development Program
at LANL.


\begin{thebibliography}{}

\bibitem[\protect\citeauthoryear{ABT}{2006}]{ABT06}
Arras P., Blaes O. M., Turner N. J., 2006, ApJ, 645, L65

\bibitem[\protect\citeauthoryear{BK}{2001}]{BK01}
Balmforth N. J., Korycansky D. G. 2001, MNRAS, 326, 833

\bibitem[\protect\citeauthoryear{BHK}{2008}]{BHK08}
Beckwith K., Hawley  J. F., Krolik  J. H., 2008, MNRAS, 390, 21

\bibitem[\protect\citeauthoryear{BHK}{2009}]{BHK09}
Beckwith K., Hawley J. F., Krolik J. H., 2009, ApJ, 707, 428

\bibitem[\protect\citeauthoryear{BSM}{2012}]{BSM12}
Belloni T. M., Sanna A.,  Mendez M., 2012, MNRAS, 426, 1701

\bibitem[\protect\citeauthoryear{Chan}{2009}]{Chan09}
Chan C.-K., 2009, ApJ, 704, 68

\bibitem[\protect\citeauthoryear{Borro}{2006}]{Borro06}
de Val-Borro M., et al., 2006, MNRAS, 370, 529

\bibitem[\protect\citeauthoryear{Villiers}{2003}]{Villiers03}
De Villiers  J.-P., Hawley  J. F., 2003, ApJ, 592, 1060

\bibitem[\protect\citeauthoryear{DGK}{2007}]{DGK12}
Dolence, J.C., Gammie, C.F., Shiokawa, H., Noble, S.C. 2012, ApJ, 746, L10



\bibitem[\protect\citeauthoryear{FBAS}{2007}]{FBAS07}
Fragile  P. C., Blaes O., Anninos P., Salmonson  J. D., 2007, ApJ, 668, 417

\bibitem[\protect\citeauthoryear{Fromang}{2007}]{Fromang07}
Fromang S., Papaloizou, J., Lesur G., Heinemann T., 2007, A\&A, 476, 1123

\bibitem[\protect\citeauthoryear{FL}{2009}]{FL09}
Fu W., Lai D., 2009, ApJ, 690, 1386

\bibitem[\protect\citeauthoryear{FL}{2011}]{FL11}
Fu W., Lai D., 2011, MNRAS, 410, 399

\bibitem[\protect\citeauthoryear{FL}{2012}]{FL12}
Fu W., Lai D., 2012, MNRAS, 423, 831

\bibitem[\protect\citeauthoryear{GS}{2005}]{GS05}
Gardiner T. A., Stone J. M., 2005, AIP Conference Proceedings, 784, 475

\bibitem[\protect\citeauthoryear{HGK}{2011}]{HGK11}
Hawley J. F., Guan X., Krolik  J. H., 2011, ApJ, 738, 84

\bibitem[\protect\citeauthoryear{HP}{2009}]{HP09}
Heinemann T., Papaloizou  J. C. B., 2009, MNRAS, 397, 52

\bibitem[\protect\citeauthoryear{HBFF}{2009}]{HBFF09}
Henisey K. B., Blaes O. M., Fragile P. C., Ferreira B. T., 2009, ApJ, 706, 705

\bibitem[\protect\citeauthoryear{HBFF}{2012}]{HBFF12}
Henisey K. B., Blaes O. M., Fragile P. C., 2012, ApJ, 761, 18

\bibitem[\protect\citeauthoryear{Horak}{2013}]{Horak13}
Horak J., Lai D. 2013, MNRAS, in press

\bibitem[\protect\citeauthoryear{Kato}{2001}]{Kato01}
Kato S., 2001, PASJ, 53, 1

\bibitem[\protect\citeauthoryear{Kato}{2003}]{Kato03a}
Kato S., 2003, PASJ, 55, 257

\bibitem[\protect\citeauthoryear{Kulkarni}{2011}]{Kulkarni11}
Kulkarni A. K., et al., 2011, MNRAS, 414, 1183

\bibitem[\protect\citeauthoryear{LT}{2009}]{LT09}
Lai D., Tsang D., 2009, MNRAS, 393, 979

\bibitem[\protect\citeauthoryear{LN04}{2004}]{LN04}
Li L., Goodman J., Narayan R., 2003, ApJ, 593, 980

\bibitem[\protect\citeauthoryear{MM}{2003}]{MM03}
Machida M., Matsumoto R., 2003, ApJ, 585, 429

\bibitem[\protect\citeauthoryear{MM}{2003}]{MM12}
McKinney J. C., Tchekhovskoy A., Blandford R. D., 2012, MNRAS, 423, 3083

\bibitem[\protect\citeauthoryear{MBM}{2007}]{MBM07}
Mignone A., Bodo G., Massaglia S., Matsakos T., Tesileanu O., Zanni C., Ferrari A., 2007, ApJS, 170, 228

\bibitem[\protect\citeauthoryear{ML}{2013}]{ML13}
Miranda R., Lai D., 2013, in preparation 

\bibitem[\protect\citeauthoryear{MG}{2009}]{MG09}
Moscibrodzka M., Gammie C. F., Dolence J. C., Shiokawa H., Leung Po Kin, 2009, ApJ, 706, 497

\bibitem[\protect\citeauthoryear{NKH}{2009}]{NKH09}
Narayan R., Goldreich P., Goodman  J., 1987, MNRAS, 228, 1

\bibitem[\protect\citeauthoryear{NKH}{2009}]{NKH09}
Noble S. C., Krolik J. H., Hawley J. F., 2009, ApJ, 692, 411

\bibitem[\protect\citeauthoryear{NKSH}{2011}]{NKSH11}
Noble S. C., Krolik J. H., Schnittman J. D., Hawley J. F., 2011, ApJ, 743, 115

\bibitem[\protect\citeauthoryear{OL}{2003}]{OL03}
Ogilvie G. I., Lubow S. H., 2003, ApJ, 587, 398

\bibitem[\protect\citeauthoryear{ORM}{2009}]{ORM09}
O'Neill S. M., Reynolds C. S., Miller C. M., 2009, ApJ, 693, 1100              

\bibitem[\protect\citeauthoryear{ORMS}{2011}]{ORMS11}
O'Neill S. M., Reynolds C. S., Miller C. M., Sorathia K., 2011, ApJ, 736, 107

\bibitem[\protect\citeauthoryear{PW}{1980}]{PW80}
Paczynski B.,  Witta P. J., 1980, A\&A, 88, 23

\bibitem[\protect\citeauthoryear{Penna}{2010}]{Penna10}
Penna R. F., McKinney J. C., Narayan R., Tchekhovskoy A., Shafee R., McClintock J. E., 2010, MNRAS, 408, 752

\bibitem[\protect\citeauthoryear{RM}{2006}]{RM06}
Remillard R. A., McClintock J. E., 2006, ARA\&A, 44, 49

\bibitem[\protect\citeauthoryear{RM}{2009}]{RM09}
Reynolds C. S., Miller M. C., 2009, ApJ, 692, 869


\bibitem[\protect\citeauthoryear{TP}{1999}]{TP99}
Tagger M., Pellat R., 1999, A\&A, 349, 1003

\bibitem[\protect\citeauthoryear{TV}{2006}]{TV06}
Tagger M., Varniere P., 2006, ApJ, 652, 1457

\bibitem[\protect\citeauthoryear{TL}{2008}]{TL08}
Tsang D., Lai D., 2008, MNRAS, 387, 446

\bibitem[\protect\citeauthoryear{Tsang}{2009}]{TL09}
Tsang D., Lai D., 2009a, MNRAS, 393, 992

\bibitem[\protect\citeauthoryear{Tsang}{2009}]{TL09}
Tsang D., Lai D., 2009b, MNRAS, 396, 589

\bibitem[\protect\citeauthoryear{Tsang}{2009}]{TL09}
Tsang D., Lai D., 2009c, MNRAS, 400, 470

\bibitem[\protect\citeauthoryear{Wagoner}{1999}]{Wagoner99}
Wagoner R. V., 2008, J. Phys: Conf. Series, Vol.118, 012006

\bibitem[\protect\citeauthoryear{YL}{2012}]{YL12}
Yu C., Lai D., 2012, MNRAS, in press
\end{thebibliography}
\end{document}